\documentstyle[11pt,aaspp]{article}
\begin{document}

\title{GRAVITATIONAL MICROLENSING IN THE LOCAL GROUP
\footnote{in press in Annual Review of Astronomy and Astrophysics, Vol. 34}}
\author{Bohdan Paczy\'nski}

\affil{Princeton University Observatory, Princeton, NJ 08544--1001}
\affil{E-mail: bp@astro.princeton.edu}


\begin{abstract}

The status of searches for gravitational microlensing events of the stars 
in our galaxy and in other galaxies of the Local Group, the interpretation 
of the results, some theory, and prospects for the future are reviewed.
The searches have already
unveiled $ \sim 100 $ events, at least two of them caused by
binaries, and have already proven to be useful for studies of the
Galactic structure.  The events detected so far are probably attributable
to the effects of ordinary stars, and possibly to sub-stellar brown
dwarfs; however a firm conclusion cannot be reached yet because
the analysis published to date is based on a total of only 16 events.
The current searches, soon to be upgraded, will probably allow determination
of the mass function of stars and brown dwarfs in the next few years;
these efforts will also
provide good statistical information about binary systems, in particular
their mass ratios.  They may also reveal the nature of dark matter and 
allow us to detect planets and planetary mass objects.

\end{abstract}

\section{Introduction}

The topic of gravitational lensing has a long history, as described
for example in the first book entirely devoted to the subject
(Schneider et al. 1992).  The first known theoretical calculation of a
light ray bending by massive objects was published by Soldner (1801),
who used Newtonian mechanics, and determined that the 
deflection angle at the solar limb should be $ 0.''84 $, half the
value calculated with the general theory of relativity (Einstein 1911,
1916).  The first observational detection of this effect came soon afterwards
(Dyson et al. 1920).  Zwicky (1937) pointed out that distant galaxies
may act as gravitational lenses.  Almost all essential formulae used today 
to analyze gravitational lensing were derived by Refsdal (1964).
The first case of a double image created by gravitational
lensing of a distant source, the quasar 0957+561, was discovered by Walsh 
et al. (1979).  Arc--like images of extended sources, the galaxies, were 
first reliably reported by Lynds \& Petrosian (1989).  In these three cases
the sun, a galaxy, and a cluster of galaxies, were acting as gravitational
lenses.  There are many recent review articles 
(Blandford \& Narayan 1992, Refsdal \& Surdej 1994, Roulet \& Mollerach 1996)
and international conferences (Moran et al. 1989, Mellier et al. 1990,
Kayser et al. 1992, Surdej et al. 1993, Kochanek \& Hewitt 1996)
on the subject of gravitational lensing.  A review of 
gravitational microlensing experiments has been published by Ansari (1995).

The effect of double imaging of a distant source by a point mass located
close to the line of sight, and acting as a gravitational lens,
has been proposed many times over.  Chang and Refsdal (1979) and
Gott (1981) noted that
even though a point mass in a halo of a distant galaxy
would create an unresolvable double image of a background quasar,
the time variation of the
combined brightness of the two images could be observed.
This way the effect of non-luminous matter
in a form of brown dwarfs or Jupiters could be detected.
The term ``microlensing'' was proposed by Paczy\'nski (1986a) to 
describe gravitational lensing which can be
detected by measuring the intensity variation of a
macro--image made of any number of unresolved micro--images.  

Paczy\'nski (1986b) suggested that a massive search
of light variability among millions of stars in the Large Magellanic
Cloud could be used to detect dark matter in the galactic halo.  Luckily,
the technology needed for such a search became available soon afterwards,
and the 1986 paper is credited with triggering the current microlensing
searches: EROS (Aubourg et al. 1993),
MACHO (Alcock et al. 1993), OGLE (Udalski 1992), and DUO (Alard 1996b,
Alard et al. 1995a).  Griest (1991) proposed that objects responsible 
for gravitational microlensing be called
massive astrophysical compact halo objects (MACHO).  The name
became very popular and it is commonly used to refer to all
objects responsible for the observed microlensing events, no matter
where they are located and what their mass may be.

It is not possible to discuss all theoretical and observational
papers related to microlensing in the Local Group within the modest
volume of this article.
The selection of references at the end of this review is
limited, and I apologize for all omissions, and for the way I made the
selection.  Let us hope that somebody will write
a more careful historical
review before too long.  Fortunately, there is a bibliography of over
one thousand papers that are related to gravitational lensing, and it 
is available electronically.  This bibliography has been compiled, and
it is continuously updated, by J. Surdej and  by A. Pospieszalska.
It can be found on the World Wide Web at:

\centerline{http://www.stsci.edu/ftp/stsci/library/grav\_ lens/grav\_ lens.html}

The MACHO and the OGLE collaborations provide up-to-date information 
about their findings and a complete bibliography of their work on the 
World Wide Web and by anonymous ftp. 
The photometry of OGLE microlensing events, their finding charts,
as well as a regularly updated OGLE status report, including more information
about the ``early warning system'', can be found over Internet from the host:

\centerline{sirius.astrouw.edu.pl   \hskip 0.5cm (148.81.8.1) ,}
\noindent
using ``anonymous ftp'' service (directory ``ogle'', files ``README'', 
``ogle.status'', ``early.warning''). The file ``ogle.status''
contains the latest news and references to all OGLE related
papers, and PostScript files of some publications.
These OGLE results are also available over World Wide Web at:

\centerline{http://www.astrouw.edu.pl \hskip 1.0cm (Europe).}
\noindent
A duplicate of this information is available at

\centerline{http://www.astro.princeton.edu/\~\/ogle/ \hskip 1.0cm (North
America).}
\noindent
A complete information about MACHO results is available at:

\centerline{http://wwwmacho.mcmaster.ca/ \hskip 1.0cm  (North America) }
\noindent
with a duplicate at

\centerline{http://wwwmacho.anu.edu.au/ \hskip 1.0cm (Australia). }
\noindent
The information about MACHO alerts is to be found at:

\centerline{http://darkstar.astro.washington.edu/ }

No doubt other groups will provide similar information before too long.
The main weakness of this electronic information distribution is the
frequent change of address and mode of access.
I shall do my best to keep a guide to this information
as part of the OGLE home page on the WWW.

In the following section a simple model of lensing by an isolated 
point mass is presented -- this is all theory one needs to understand most
individual microlensing events.  The third section presents a
model for the space distribution and kinematics of lensing objects
in order to provide some insight into problems in relating the observed
time scales of microlensing events to the masses of lensing objects.
The fourth section provides a glimpse of diversity of light curves
due to lensing by double objects, like binary stars or planetary systems.
Some of the special effects which make microlensing more complicated
than originally envisioned are described in the fifth section.
The most essential information about the current searches for 
microlensing events, and some of the results as well as problems with 
some results, are presented in section six.  The last section is a rather
personal outline of the prospects for the future of microlensing searches.

\section{Single point mass lens}

Let us consider a single point mass $ M $ (a deflector)
at a distance $ D_d $ from the observer, and a point source $ S $
at a distance $ D_s $ from the observer, as in Figure 1.  Let there
be two planes perpendicular to the line of sight, at the deflector
and the source distances, respectively.  The deflector  has
angular coordinates $ (x_m,y_m) $ in the sky, as seen by the
observer.  This projects into points $ M $ and $ M_s $
in the two planes, with the corresponding linear coordinates:

\begin{figure}[p]
\vspace{8cm}
\includegraphics{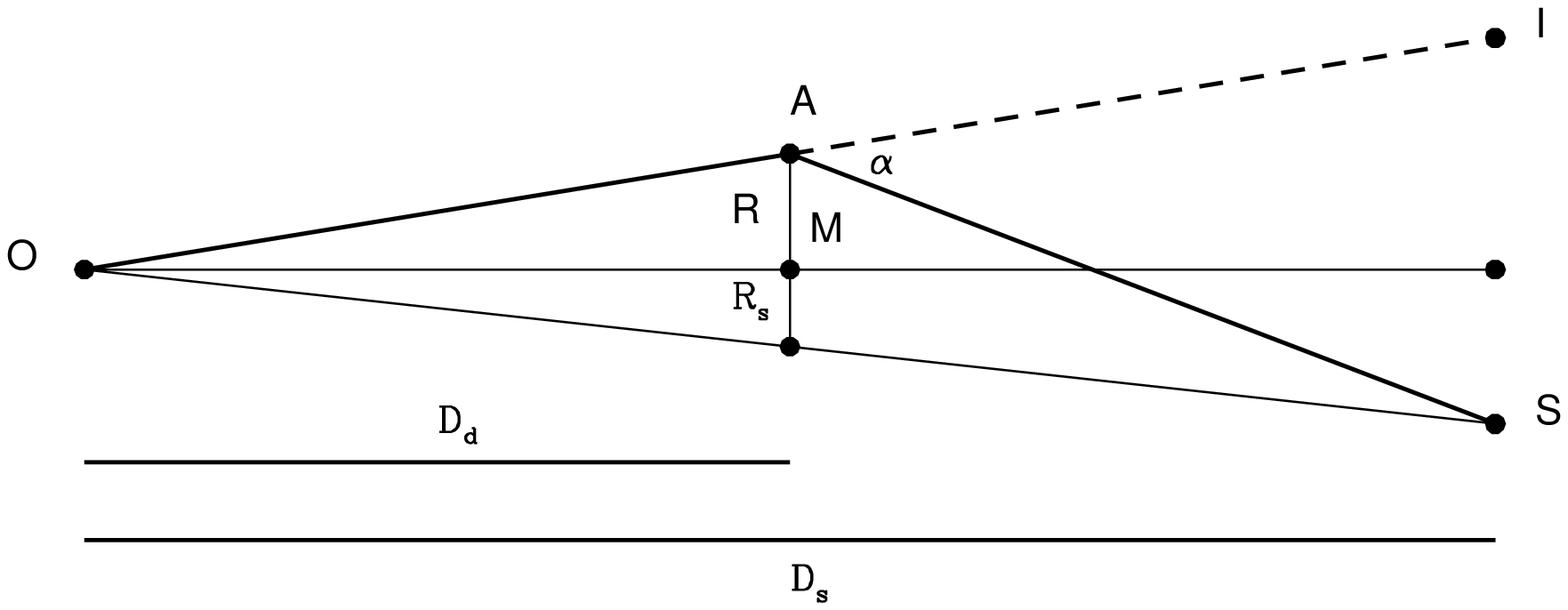}
\caption{\small
The geometry of gravitational lensing is shown.  The observer,
the lensing mass, and the source are located at the points $ O $,
$ M $, and $ S $, respectively.  The light rays are
deflected near the lensing mass by the angle $ \alpha $,
and the image of the source appears to be located at the point $ I $,
not at $ S $.  The distances from the observer to the lens
(deflector) and to the source are indicated as $ D_d $ and $ D_s $,
respectively.
}
\vspace{8.5cm}
\includegraphics{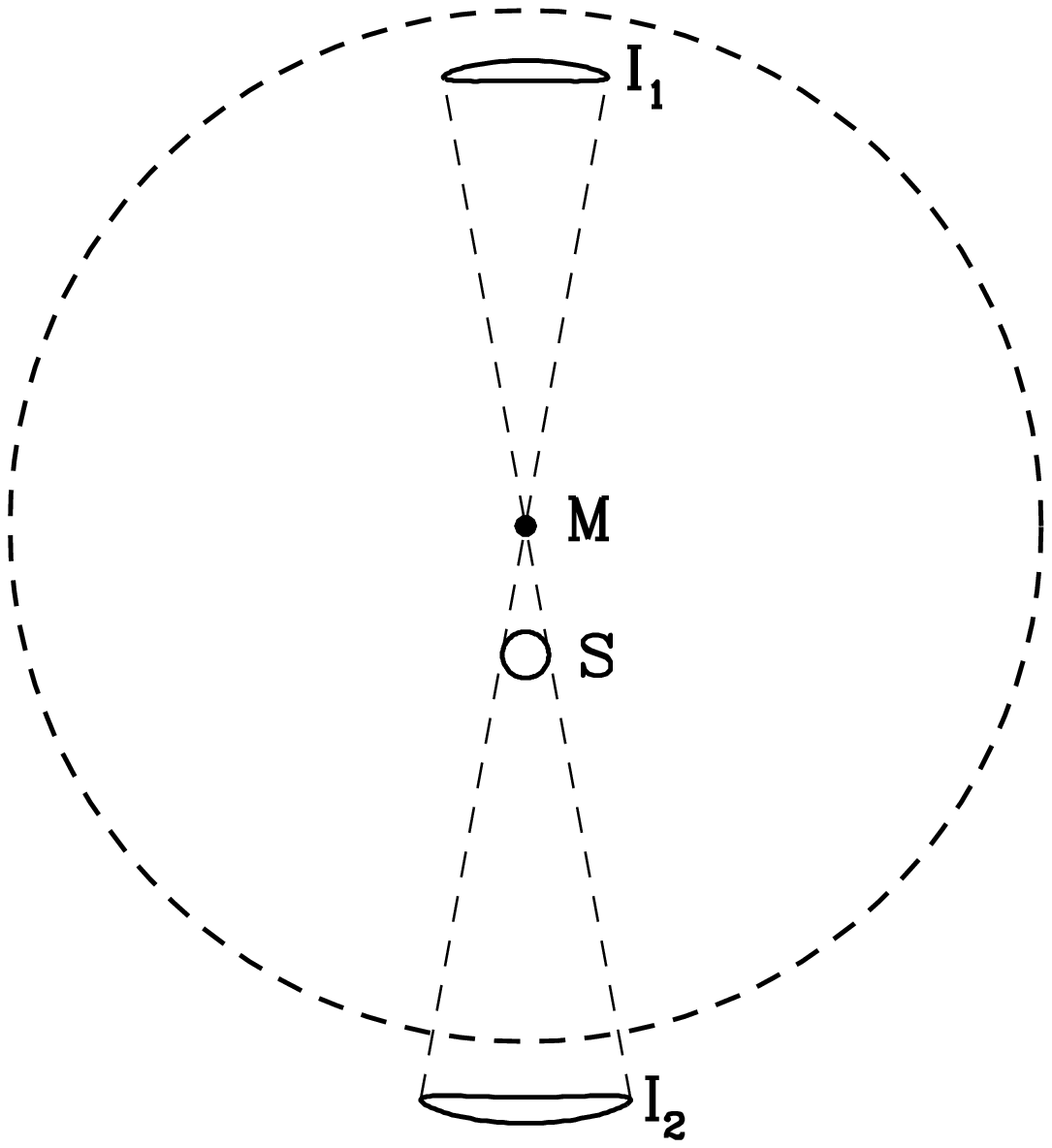}
\caption{\small
The geometry of gravitational lensing is shown.  The
lensing mass, the small circular source, and the two images are marked with
$ M $, $ S $, $ I_1 $, and $ I_2 $, respectively.  Of course, in the
presence of mass $ M $ the source is not seen at $ S $ but only at
$ I_1 $ and $ I_2 $.  The Einstein 
ring is shown as a dashed circle.  A typical radius of the circle
is $ \sim 1 $ milli arc second for microlensing by stars in our galaxy.
}
\end{figure}

$$
X_M = x_m D_d , \hskip 1.0cm
Y_M = y_m D_d , \eqno(1a)
$$
$$
X_{M,s} = x_m D_s , \hskip 1.0cm
Y_{M,s} = y_m D_s . \eqno(1b)
$$

Let the observer, located at point $ O $, look at the sky in the direction
with angular coordinates $ (x,y) $.  The line of sight intersects
the deflector plane at point $ A $ with coordinates:
$$
X_A = x D_d , \hskip 1.0cm
Y_A = y D_d . \eqno(2a)
$$
If there was no deflection of light by the massive object then the line of 
sight would intersect the source plane at point $ I $ with coordinates:
$$
X_I = x D_s , \hskip 1.0cm
Y_I = y D_s . \eqno(2b)
$$
In fact the light ray passes the deflector at a distance 
$$
R = \left[ \left( X_A - X_M \right) ^2 +
\left( Y_A - Y_M \right) ^2 \right] ^{1/2} .  \eqno(3)
$$
As a consequence of general relativity the light ray is deflected
by the angle
$$
\alpha = { 4GM \over R c^2 } , \eqno(4)
$$
with the two components:
$$
\alpha _x = \alpha { X_A - X_M \over R } , \hskip 1.0cm
\alpha _y = \alpha { Y_A - Y_M \over R } .  \eqno(5)
$$
The deflected light ray intersects the source
plane at the point $ S $ with the coordinates:
$$
X_S = X_I - \alpha _x (D_s - D_d) ,  \hskip 1.0cm
X_S = Y_I - \alpha _y (D_s - D_d) ,  \eqno(6)
$$
Passing through three points: $ O $, $ A $ , and $ S $, the light ray 
defines a plane.  The lensing mass $ M $, and the point $ I $ 
are also located in the same plane, as shown in Fig. 1.  If there was no
effect of the mass $ M $ on the light rays then the source of light would
be seen at the point $ S $, at the angular distance $ R_s/D_d $
from the point $ M $.  However, as the light rays are deflected by
$ M $ the image of the source appears not at S but at the point $ I $, 
at the angular distance $ R/D_d $ from the point $ M $.  It is clear
that the distance $ (R+R_s) $ in the deflector plane is proportional
to the distance between $ I $ and $ S $ in the source plane, and the
latter can be calculated with the eqs. (6).  Combining this
with the eqs. (5) and (4) we obtain:
$$
R + R_s = 
\left[ \left( X_S - X_I \right) ^2 +
\left( Y_S - Y_I \right) ^2 \right] ^{1/2}
{ D_d \over D_s } =
\alpha \left( D_s - D_d \right) { D_d \over D_s } =
{ 4 GM \over R c^2 } { \left( D_s - D_d \right) D_d \over D_s } .  \eqno(7)
$$
This equation may be written as
$$
{ R_{\rm s} \over R_{\rm E} } = 
- { R \over R_{\rm E} } + { R_{\rm E} \over R },
\hskip 1.0cm R^2 + R_{\rm s} R - R_{\rm E}^2 = 0 ,  \eqno(8a)
$$
where
$$
R_E^2 \equiv 2 R_g D , \hskip 1.0cm
R_g \equiv { 2GM \over c^2 } , \hskip 1.0cm
D \equiv { \left( D_s - D_d \right) D_d \over D_s } , \eqno(8b)
$$
and $ R_E $, $ R_g $, and $ D $ are called the linear Einstein ring
radius of the lens, the gravitational radius of the mass $ M $, and 
the effective lens distance, respectively.

The equation (8a) has two solutions:
$$
R_{+,-} = 0.5 \left[ R_s \pm \left( R_s^2 + 4 R_E^2 \right) ^{1/2} \right] .
\eqno(9)
$$
These two solutions correspond to the two images of the same source,
located on the opposite sides of the point $ M $, at the angular distances
of $ R_+/D_d $ and $ R_-/D_d $, respectively.  The appearance of a small
circular source (in the absence of lensing)
and its two distorted two images in a telescope with very high resolving
power is shown in Fig. 2.  As any lensing conserves surface brightness
(cf. Schneider et al. 1992, section 5.2),
the ratio of the image to source intensity is given by the ratio of
their areas.  With the geometry shown in Fig. 2 this ratio can be
calculated as
$$
A_{+,-} = \left| { R_{+,-} \over R_s } ~ { d R_{+,-} \over d R_s } \right|
= { u^2 + 2 \over 2 u \left( u^2 + 4 \right) ^{1/2} } \pm 0.5 ,
\hskip 1.0cm u \equiv { R_s \over R_E } , \eqno(10)
$$
assuming that the source is very small.  The quantity $ A $ is
called magnification or amplification.  The term magnification
will be used throughout this paper, as it better describes the
lensing process.  The total magnification of the two images can
be calculated as
$$
A = A_+ + A_- = { u^2 + 2 \over u \left( u^2 + 4 \right) ^{1/2} } . \eqno(11)
$$
It is interesting to note that this is always larger than unity.
It is also interesting that the difference in the magnification
of the two images is constant:
$$
A_+ - A_- = 1.    \eqno(12)
$$

Let us consider a typical galactic case, with a lens of
$ \sim 1 ~ M_{\odot} $ mass at a distance of a few kiloparsecs, 
and the source at a larger
distance.  The angular Einstein ring radius, $ r_{_E} $,
can be calculated with the eqs. (8b) to be
$$
r_{_E} \equiv { R_E \over D_d } = 
\left[ \left( { 4GM \over c^2 } \right)
~ \left( { D_s - D_d \over D_s D_d } \right) \right] ^{1/2} = 
$$
$$
= 0.902 ~ mas ~
\left( { M \over M_{\odot} } \right) ^{1/2}
\left( { 10 ~ kpc \over D_d } \right) ^{1/2}
\left( 1 - { D_d \over D_s } \right) ^{1/2} .  \eqno(13)
$$
With the image separation $ \sim 2 r_{_E} $, i.e.
of the order of a milliarcsecond (mas), we can only see the combined light
intensity, rather than two separate images because of the limited
resolution of optical telescopes.  Fortunately, all objects
in the galaxy move, and we may expect a relative proper motion to be
$$
\dot r = { V \over D_d } = 4.22 ~ mas ~ yr^{-1} ~ 
\left( { V \over 200 ~ km ~ s^{-1} } \right)
\left( { 10 ~ kpc \over D_d } \right) ,  \eqno(14)
$$

\begin{figure}[t]
\vspace{8cm}
\includegraphics{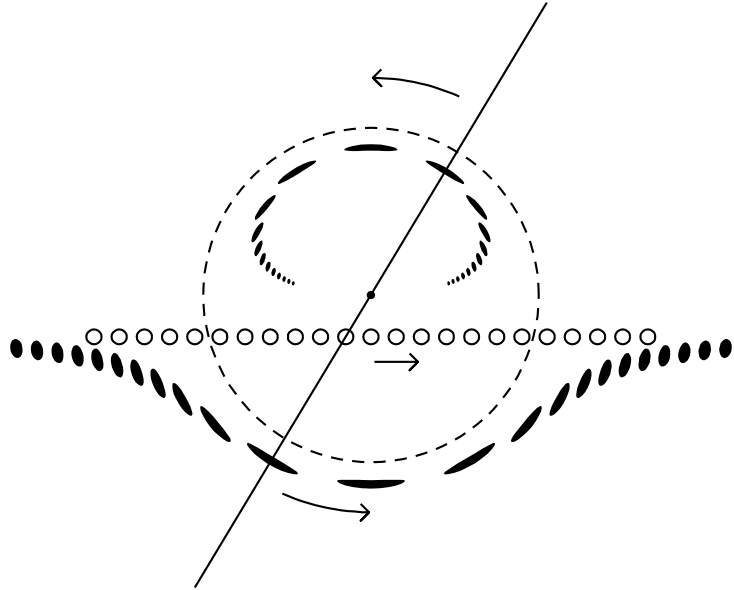}
\caption{\small
The geometry of gravitational lensing is shown.  
The lensing mass is indicated with a dot at the center
of the Einstein ring, which is marked with a dashed line.
The source positions are shown with a series of small open circles.
The locations and the shapes of the two images are shown with a series of
dark ellipses.  At any instant the two images, the source and the lens
are all on a single line, as shown in the figure for one particular instant.
}
\end{figure}

\noindent
where $ V $ is the relative transverse velocity of the lens with
respect to the source.  Combining the last two equations we can
calculate the characteristic time scale for a microlensing phenomenon
as the time it takes the source to move with respect to the lens by
one Einstein ring radius:
$$
t_0 \equiv { r_{_E} \over \dot r } =
0.214 ~ yr ~ 
\left( { M \over M_{\odot} } \right) ^{1/2}
\left( { D_d \over 10 ~ kpc } \right) ^{1/2}
\left( 1 - { D_d \over D_s } \right) ^{1/2} 
\left( { 200 ~ km ~ s^{-1} \over V } \right) .  \eqno(15)
$$
This definition is almost universally accepted, with one major exception:
the MACHO collaboration multiplies the value of $ t_0 $ as given
with the eq. (15) by a factor 2.

\begin{figure}[p]
\vspace{8.5cm}
\includegraphics{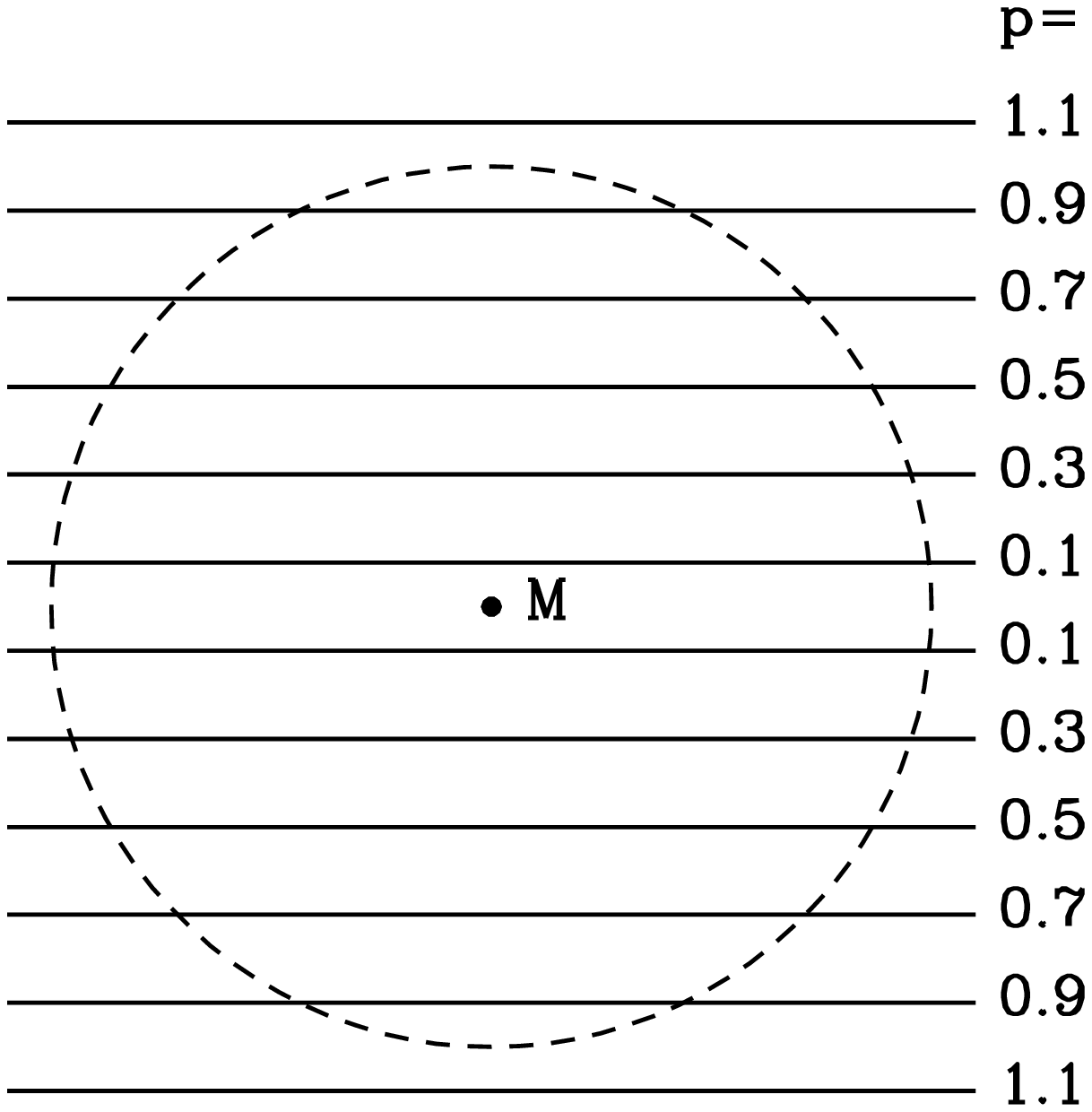}
\caption{\small
The geometry of gravitational lensing is shown. 
The lensing mass $ M $ is located at the center
of the Einstein ring, which is marked with a dashed line.
The twelve horizontal lines represent relative trajectories of the
source, labeled with the value of dimensionless
impact parameter $ p $.
}
\vspace{8cm}
\includegraphics{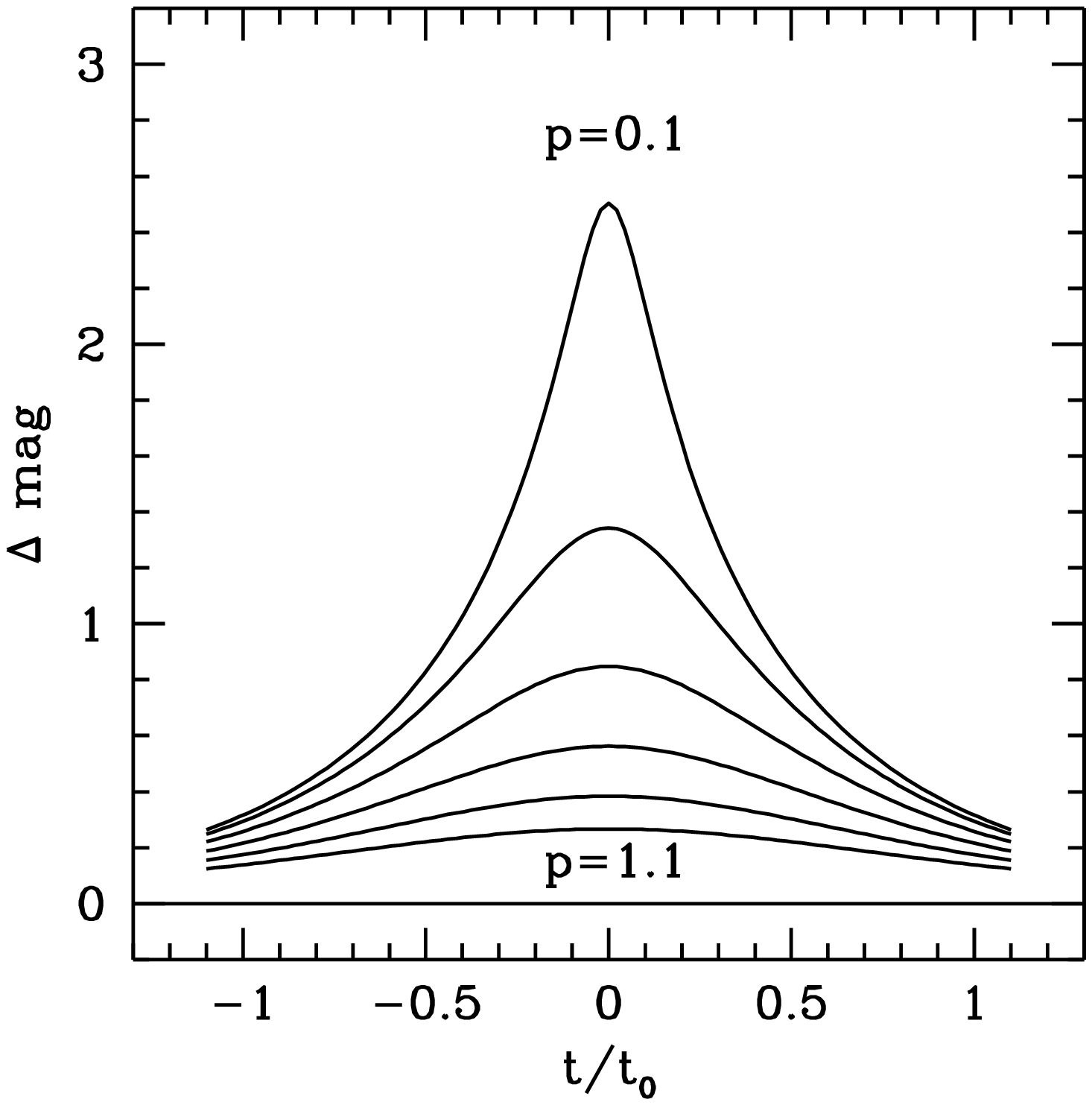}
\caption{\small
The variation of the magnification due to a point gravitational
lensing is shown in stellar magnitudes as
a function of time.  The unit $ t_0 $ is defined
as the time it takes the source to move a distance equal to the
Einstein ring radius, $ r_E $.  The six light curves correspond
to the six values of the dimensionless impact parameter:
$ p = 0.1, ~ 0.3, ~ 0.5, ~ 0.7, ~ 0.9, ~ 1.1 $.
}
\end{figure}

While the lens moves with respect to the source the two images change
their position and brightness, as shown in Fig. 3.  When
the source is close to the lens the images are highly elongated and
their proper motion is much higher than that of the source.
Note, that for each source position
the two images, the source and the lens are all
located on a straight line which rotates around the lens.
Unfortunately, none of this geometry can be
observed directly for the stellar mass lenses, as the angular scale
is of the order of a milli arcsecond.  The total light variations
can be observed, and the light curve can be calculated
with the eq. (11).  The variable $ u $ can be determined as
$$
u = \left[ p^2 + \left( { t-t_{max} \over t_0 } \right) ^2 \right] ^{1/2} ,
\eqno(16)
$$
where $ p $ is the dimensionless impact parameter: the smallest angular
distance between the source and the lens measured in units of Einstein
ring radius, and $ t_{max} $ is the time of the maximum magnification
by the lens.  The geometry of lensing for six values of impact
parameter, and the corresponding time variability expressed in stellar
magnitudes: $ \Delta m \equiv 2.5 \log A $, are shown in Figs. 4 and 5.

It is customary to define the cross section for gravitational microlensing
to be equal to the area of the Einstein circle.  The combined magnification
of the two images of a source located within that circle is larger than
$$
A \geq { 3 \over 5^{1/2} } = 1.3416 , \hskip 1.0cm
\Delta m  \geq 0.3191 ~ mag  ,  \hskip 1.0cm
( u \leq 1 )  ,  \eqno(17)
$$
(cf. eq. 11) i.e. it is easy to detect such intensity variation
with a reasonably accurate photometry.

Let there be many lensing objects in the sky.  The fraction of solid
angle covered with their Einstein rings is called the optical depth
to gravitational microlensing.  Let
all lensing objects have identical masses $ M $.
In a thin slab at a distance $ D_d $ and a thickness $ \Delta D_d $,
there is, on average, one lens per surface area
$ \pi R_M^2 = M /( \rho \Delta D_d ) $, where $ \rho $ is the average
mass density due to lenses in the volume
$ \pi R_{\rm M}^2 \Delta D_{\rm d} $.  Each lens has
a cross section $ \pi R_E^2 $, with the Einstein ring radius $ R_E $ given
with the eqs. (8b).  The slab contribution to the optical
depth is given as
$$
\Delta \tau = { \pi R_E^2 \over \pi R_M^2 } =
\left[ { 4 \pi G \rho \over c^2 } ~
{ D_d \left( D_s - D_d \right) \over D_s } \right] \Delta D_d .  
\eqno(18)
$$
The total optical depth due to all lenses between the source and the
observer can be calculated as
$$
\tau = \int _0^{D_s} 
{ 4 \pi G \rho \over c^2 } ~ { D_d \left( D_s - D_d \right) \over D_s } ~
d D_d =
{ 4 \pi G \over c^2 } D_s^2 \int _0^1 \rho (x) ~ x(1-x) ~ dx ,
\eqno(19)
$$
where $ x \equiv { D_d / D_s } $.  Note that the optical depth $ \tau $
depends on the total mass in all lenses, but it is independent of the
masses of individual lenses, $ M $.
If the density of matter is constant then we have
$$
\tau = { 2 \pi \over 3 } ~ { G \rho \over c^2 } ~ D_s^2 .  \eqno(20)
$$

If the system of lenses is self-gravitating, then a crude but very simple 
estimate of the optical depth can be made.  Let us suppose that the 
distance to the source $ D_s $
is approximately equal to the size of of the whole system, a galaxy of
lenses.  The virial theorem provides a relation
between the velocity dispersion $ V^2 $, the density $ \rho $, and the 
size $ D_s $:
$$
{GM_{tot} \over D_s } \approx
{ G \rho D_s^3 \over D_s } \approx V^2 .  \eqno(21)
$$
Combining eqs. (20) and (21) we obtain
$$
\tau \approx { V^2 \over c^2 } .   \eqno(22)
$$ 
A more accurate estimate of the optical depth can be obtained by
evaluating the integral in eq. (19) for any distribution of mass
density along the line of sight.


\section{The event rate and the lens masses}

We now proceed to the estimate of the number of microlensing events $ N $
that may be expected if $ n $ sources are monitored over a time interval
$ \Delta t $.  
We consider only those microlensing events which have peak magnification
in excess of $ 3/ \sqrt{5} $, i.e. their dimensionless impact parameters are
smaller than unity (cf. Figures 4 and 5).  
We begin with the simplest case: all lensing objects
have the same mass $ M $,
and all have the same 3-dimensional velocity $ V $.  We also assume that
the velocity vectors have a random but isotropic distribution, the
source located at the distance $ D_s $ is stationary, and the
number density of lensing objects is statistically uniform between 
the observer and the source.

The time scale of a microlensing event is given as
$$
t_0 = { r_{_E} \over \dot r } =
{ R_E \over V_{t} } = { R_E \over V \sin i } , \eqno(23)
$$
(cf. eq. 15)
where $ i $ is the angle between the velocity vector and the line of
sight, and $ V_t = V \sin i $ is the transverse velocity of the lens.
If all events had identical time scales, then the number of microlensing
events expected in a time interval $ \Delta t $ would be given as
$$
N = { 2 \over \pi } ~ n \tau ~ { \Delta t \over t_0 } ,  
\eqno(24)
$$
for $ t_0 = const $,
where $ 2 / \pi $ is the ratio of Einstein ring diameter to its area, in
dimensionless units, and $ \tau $ is the optical depth.

In fact there is a broad distribution of event time scales as lenses
(all with the same mass $ M $ and the same space velocity $ V $) have
transverse velocities in the range $ 0 \leq V_t \leq V $, and
distances in the range $ 0 \leq D_d \leq D_s $.  Straightforward but
tedious algebra leads to the equation
$$
N = { 3 \pi \over 16 } ~ n \tau ~ { \Delta t \over t_{m} } =
\int _0^{\infty} N'(t_0) ~ dt_0 ,
\eqno(25)
$$
where
$$
t_{m} \equiv
\left( { R_E \over V } \right) _{D_d=0.5 D_s} =
\left( { GMD_s \over c^2 } \right) ^{1/2} { 1 \over V } ,  \eqno(26)
$$
is the time scale for a microlensing event due to a lens located
half way between the source and the observer, and moving with the
transverse velocity $ V $.  A detailed analysis is given by Mao \&
Paczy\'nski (1996).

The probability distribution of event time scales is
very broad.  It can be shown to have power-law tails for very short and 
for very long time scales:
$$
P(t_0 \leq t' ) = { 128 \over 45 \pi ^2 } \left( t' \over t_m \right) ^3 ,
\hskip 1.0cm {\rm for} \hskip 0.5cm t' \ll t_m ,  
\eqno(27)
$$
$$
P(t_0 \geq t' ) = { 128 \over 45 \pi ^2 } \left( t_m \over t' \right) ^3 ,
\hskip 1.0cm {\rm for} \hskip 0.5cm t' \gg t_m .
\eqno(28)
$$
Note that the power law tails in the distribution of event time scales
are generic to almost all lens distributions
ever proposed.  The very short events are due to lenses which
are either very close to the source or very close to the observer, while
the very long events are caused by the lenses which move almost along
the line of sight.  Also note that
only the first two moments of the distribution are finite;
the third and higher moments diverge.
Therefore, it is convenient to use a logarithmic probability 
distribution defined as
$$
p( \log t_0 ) ~ d \log t_0 =
\left( { \ln 10 \over N } \right) ~ 
t_0 ~ N'(t_0) ~ d \log t_0 , \eqno(29)
$$
as all moments of this distribution are finite.  

De R\'ujula et al. (1991)
proposed to use the moment analysis to deduce the distribution of
lens masses.  We shall carry on such an analysis, but first we have 
to make our model more realistic.
As most astrophysical objects have a broad range of velocities,
we adopt a 3-dimensional gaussian distribution:
$$
p (V) ~ { dV \over V_{rms} } = 3 \left( 6 \over \pi \right) ^{1/2} ~ 
\exp \left( - { 3 V^2 \over 2 V_{rms}^2 } \right) ~ 
{ V^2 ~ \over V_{rms}^2 } ~ { dV \over V_{rms} } ,
\hskip 1.0cm {\rm for} \hskip 0.5cm 0 \leq V < \infty .  
\eqno(30)
$$
where the 3--dimensional rms velocity $ V_{rms} $ is defined as
$$
V_{rms}^2 \equiv \int _0^{\infty} V^2 p (V) ~ { dV \over V_{rms} } ,  \eqno(31)
$$
We also adopt a general power law distribution of the number of 
lensing masses:
$$
n (M) ~ dM \sim M^{ \alpha } ~ dM ,
\hskip 1.0cm {\rm for} \hskip 0.5cm M_{min} \leq M \leq M_{max} ,
\eqno(32)
$$
or, in a logarithmic form:
$$
n ( \log M ) ~ d \log M = \left[ { ( \alpha +1 ) \ln 10 \over 
M_{max}^{ \alpha +1} - M_{min}^{ \alpha +1} } \right] ~
M^{ \alpha + 1 } ~ d \log M ,
\hskip 1.0cm {\rm for} \hskip 0.5cm \alpha \neq -1 ,
\eqno(33)
$$
$$
n ( \log M ) ~ d \log M = \left[ { \ln 10 \over 
\ln \left( M_{max} / M_{min} \right) } \right] ~
d \log M ,
\hskip 1.0cm {\rm for} \hskip 0.5cm \alpha = -1 .
\eqno(34)
$$
The case $ \alpha = -2 $ corresponds to equal total mass per
decade of lens masses, i.e. each decade has the same contribution
to the optical depth.  The case $ \alpha = -1.5 $ corresponds to
equal rate of microlensing events per decade of lens masses.
If the mass function is very broad, i.e. $ M_{max}/M_{min} \gg 1 $,
then the distribution of event rate will be flat in this case, with
equal number of events per logarithmic interval of $ t_0 $, with
power law tails (cf. eqs. 27 and 28).  If $ \alpha = -1 $ then
there is equal number of lenses per logarithmic interval of their masses.

A characteristic mass scale for the lenses can be related to the first
moment of the $ t_0 $ distribution:
$$
M_0 \equiv { c^2 V_{rms}^2 \over G D_s } ~ t_{0,av}^2 ,
\hskip 1.0cm {\rm where} \hskip 0.5cm
\log t_{0,av} \equiv \langle \log t_0 ~ \rangle  ,  \eqno(35)
$$
(cf. eqs. 26, 30).
The value of $ M_0 $ is indicative of the most common lens mass.
However, for $ \alpha \ll -1.5 $ the mass spectrum
is dominated by low mass objects, the event time scales
have a small effective range, and the events are mostly due to
lenses with $ M \approx M_{min} $.  In the opposite case, with
$ \alpha \gg -1.5 $, the microlensing events are dominated by lenses
with $ M \approx M_{max} $.

For a given mass range the standard deviation of $ \log t_0 $ is the
largest for $ \alpha = -1.5 $, and it can be estimated analytically.
First, imagine that there is a unique relation
between the lens mass $ M $ and the event time scale $ t_0 $.  The
exponent $ \alpha = -1.5 $ implies a uniform distribution of event
time scales in $ \log t_0 $ over the range 
$ \log t_{min} \leq \log t_0 \leq \log t_{max}$, with
$ \log ~ (t_{max}/t_{min}) = 0.5 \log ~ (M_{max}/M_{min}) = 
0.5 \Delta \log M $.  This distribution would have a standard deviation
$ \sigma _{ \log t_0} = ( \Delta \log M )^2/48 $.
However, there is no unique relation between
the lens mass and the corresponding time scale $ t_0 $ in our model.
Even if all lenses had the same mass the standard deviation
would be $ \sigma _0 = 0.268 $, according to numerical calculations.
Therefore, the following formula can be used to calculate the second moment:
$$
\sigma _{ \log t_0 } = 
\left[ \sigma _0^2 + { ( \Delta \log M )^2 \over 48 } \right] ^{1/2} ,
\hskip 0.5cm \sigma _0 = 0.268 ,
\hskip 1.0cm {\rm for} \hskip 0.5cm \alpha = -1.5 .
\eqno(36)
$$

There is another serious problem affecting the relation between
the observed event time scales and the lens masses. In general
we do not know which model should be adopted for the space distribution of
lenses and for their
kinematics.  For the purpose of our exercise we adopted a uniform space 
density and a uniform velocity distribution, all the way from the source 
to the observer.  Unfortunately, there is no consensus yet
on the distribution and the kinematics of the
dominant lensing component towards the galactic bulge and towards the LMC.
To reach a consensus it will be necessary to determine observationally
the variation of optical depth to microlensing with the galactic coordinates,
and to use this information to determine the space distribution of
the dominant lens population.  Next, a model of the galactic gravitational
potential will have to be used to estimate the
kinematics of the dominant lens population.  In order to have
confidence that the observed distribution of event time scales
is not truncated by instrumental effects the power law tails,
like those given with the eqs. (27) and (28), will have to be 
well sampled.  When all this work is done 
a relation between the distribution of event
time scales and the distribution of lens masses will be sound.

\pagebreak
\section{Double lenses and planets}

Let us consider now a large number of point mass lenses, all located 
at the same distance $ D_d $, in front of a point source at $ D_s $.  Let 
a lens number ``i'' has a mass $ M_i $ located at $ (X_i, Y_i) $.
A light ray crossing the deflector's plane at $ (X,Y) $ would pass
at a distance $ R_i $ from the lens ``i'':
$$
R_i = \left[ \left( X - X_i \right) ^2 +
\left( Y - Y_i \right) ^2 \right] ^{1/2} ,  \eqno(37)
$$
(cf. eq. 3).  The contribution to the light ray deflection by the mass ``i''
is given by the angle
$$
\alpha _i = { 4GM_i \over R_i c^2 } , \eqno(38)
$$
with the two components:
$$
\alpha _{x,i} = \alpha _i ~ { X - X_i \over R_i } , \hskip 1.0cm
\alpha _{y,i} = \alpha _i ~ { Y - Y_i \over R_i } ,  \eqno(39)
$$
(cf. eqs. 4 and 5).
As a result the combined deflection by all lensing masses the light
ray will cross the source plane at $ (X_S,Y_S) $:
$$
X_S = X { D_s \over D_d } - \sum _i \alpha _{x,i} (D_s - D_d) ,  
\hskip 1.0cm 
Y_S = Y { D_s \over D_d } - \sum _i \alpha _{y,i} (D_s - D_d) ,  \eqno(40)
$$
(cf. eq. 6).   Note that $ X, ~ Y, ~ X_i, ~ Y_i, $ and $ R_i $ are all
measured in the deflector plane, while $ X_s $ and $ Y_s $ are measured
in the source plane.

Combining eqs. (38--40) we obtain
$$
x_s = x - { (D_s - D_d) \over D_s D_d } ~
\sum _i { 4 GM_i \over c^2 } ~ { (x - x_i) \over r_i^2 } , \eqno(41)
$$
$$
y_s = y - { (D_s - D_d) \over D_s D_d } ~
\sum _i { 4 GM_i \over c^2 } ~ { (y - y_i) \over r_i^2 } , \eqno(42)
$$
where the angles are defined as
$$
x_s = { X_S \over D_s } , \hskip 0.5cm
y_s = { Y_S \over D_s } , \eqno(43)
$$
and
$$
x = { X \over D_d } , \hskip 0.5cm
y = { Y \over D_d } , \hskip 0.5cm
x_i = { X_i \over D_d } , \hskip 0.5cm
y_i = { Y_i \over D_d } , \hskip 0.5cm
r_i = { R_i \over D_d } . \eqno(44)
$$
With dimensionless masses $ m_i $ defined as
$$
m_i = { (D_s - D_d) \over D_s D_d } ~ { 4 GM_i \over c^2 } , \eqno(45)
$$
the eqs. (41--42) may be written as
$$
x_s = x - ~ \sum _i ~ { m_i (x - x_i) \over r_i^2 } , \eqno(46)
$$
$$
y_s = y - ~ \sum _i ~ { m_i (y - y_i) \over r_i^2 } . \eqno(47)
$$
All these are angles as seen by the observer.

The set of equations (46--47) can be used to find all images created
by a planar lens system made of any number of point masses
(cf. Witt 1993, and references therein).  Here we shall
restrict ourselves to a double lens case, as binary stars are known
to be very common (Abt 1983, Mao \& Paczy\'nski 1991).  A very thorough
analysis of double star microlensing is provided by Schneider \& Weiss 
(1986).  A major new phenomenon not present in a single lens case is 
the formation of caustics caused by the lens astigmatism.  When a source
crosses a caustic a new pair of images forms or disappears.  A point source
placed at a caustic is magnified by an infinite factor, while a source 
with a finite size is subject to a large, but finite magnification.
A double lens is vastly more complicated than a single one.

The eqs. (46--47) applied to the binary case are:
$$
x_s = x - ~ { m_1 (x - x_1) \over r_1^2 } -
{ m_2 (x - x_2) \over r_2^2 } , \eqno(48)
$$
$$
y_s = y - ~ { m_1 (y - y_1) \over r_1^2 } 
- ~ { m_2 (y - y_2) \over r_2^2 } . \eqno(49)
$$
It is customary to adopt $ m_1 + m_2 = 1 $.  This makes
all the angles expressed in units of the Einstein ring radius
for a lens with a unit mass.  If the binary orbital motion
is neglected (a `static binary' case) then, in addition to all standard
parameters describing single point mass lensing, there are
three new dimensionless parameters: the mass ratio: 
$ m_1 / m_2 $, the binary separation in units of the Einstein ring radius,
and the angle between the source trajectory
and the line joining the two components of the lens.  The diversity of
possible light curves is staggering.  The first computer code that can 
not only generate theoretical light curves for a binary lens, but can also
fit the best theoretical light curve to the actual data, was developed by
Mao \& Di Stefano (1995).  It was applied to determine the parameters of
two events which were almost certainly caused by double lenses:
OGLE \#7 (Udalski et al. 1994d), and DUO \#2 (Alard et al. 1995b).

\begin{figure}[p]
\vspace{8cm}
\includegraphics{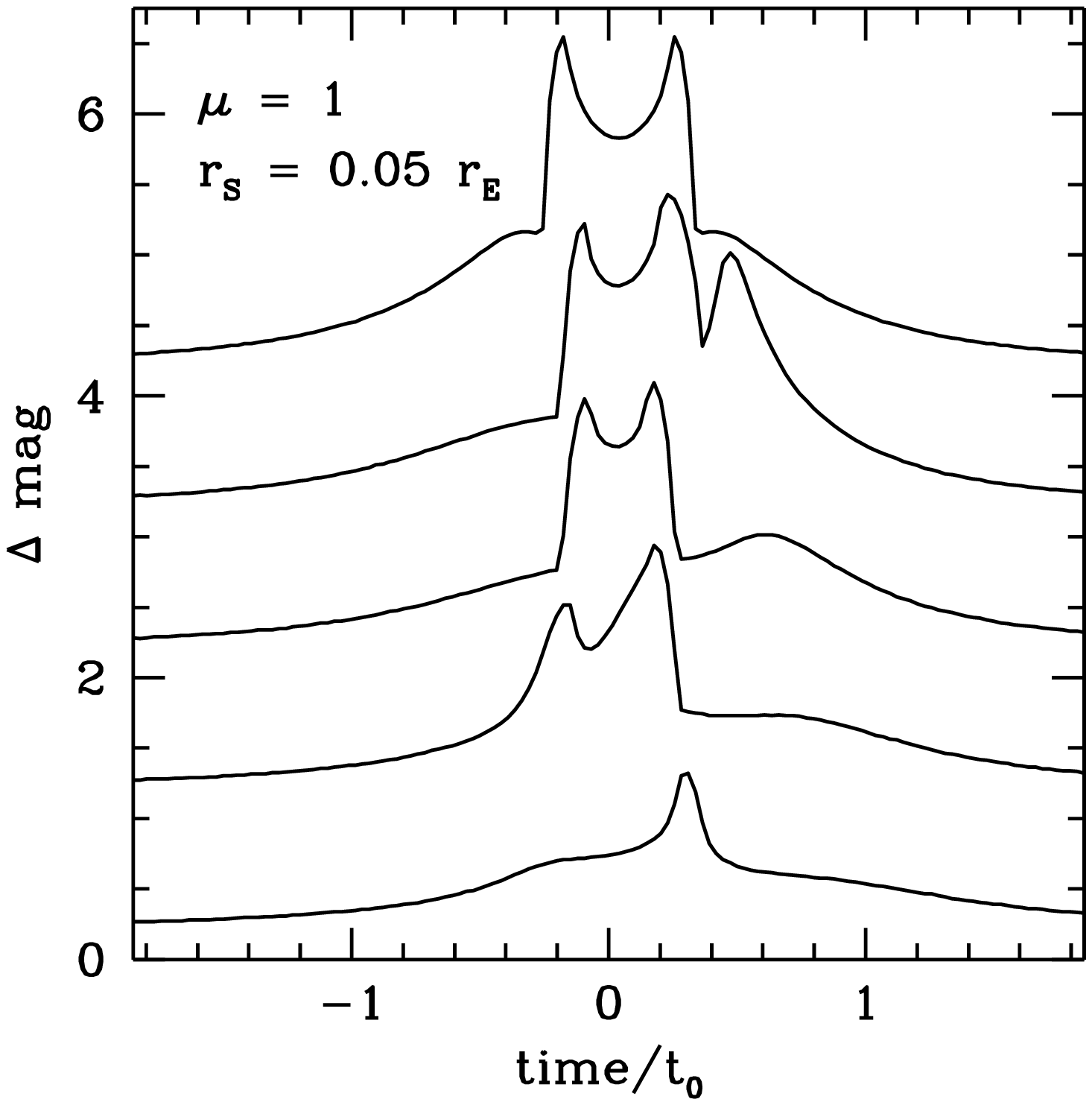}
\caption{\small
Five light curves are shown as examples of binary microlensing.
The two components of the binary have identical masses and are separated
by one Einstein ring radius.  The corresponding source trajectories
are shown in Figure 7.  The top light curve shown here corresponds
to the top trajectory in Figure 7.
The sharp spikes are due to caustic crossings
by the source.  (The light curves are shifted by one magnitude for clarity
of the display).
}
\vspace{7cm}
\includegraphics{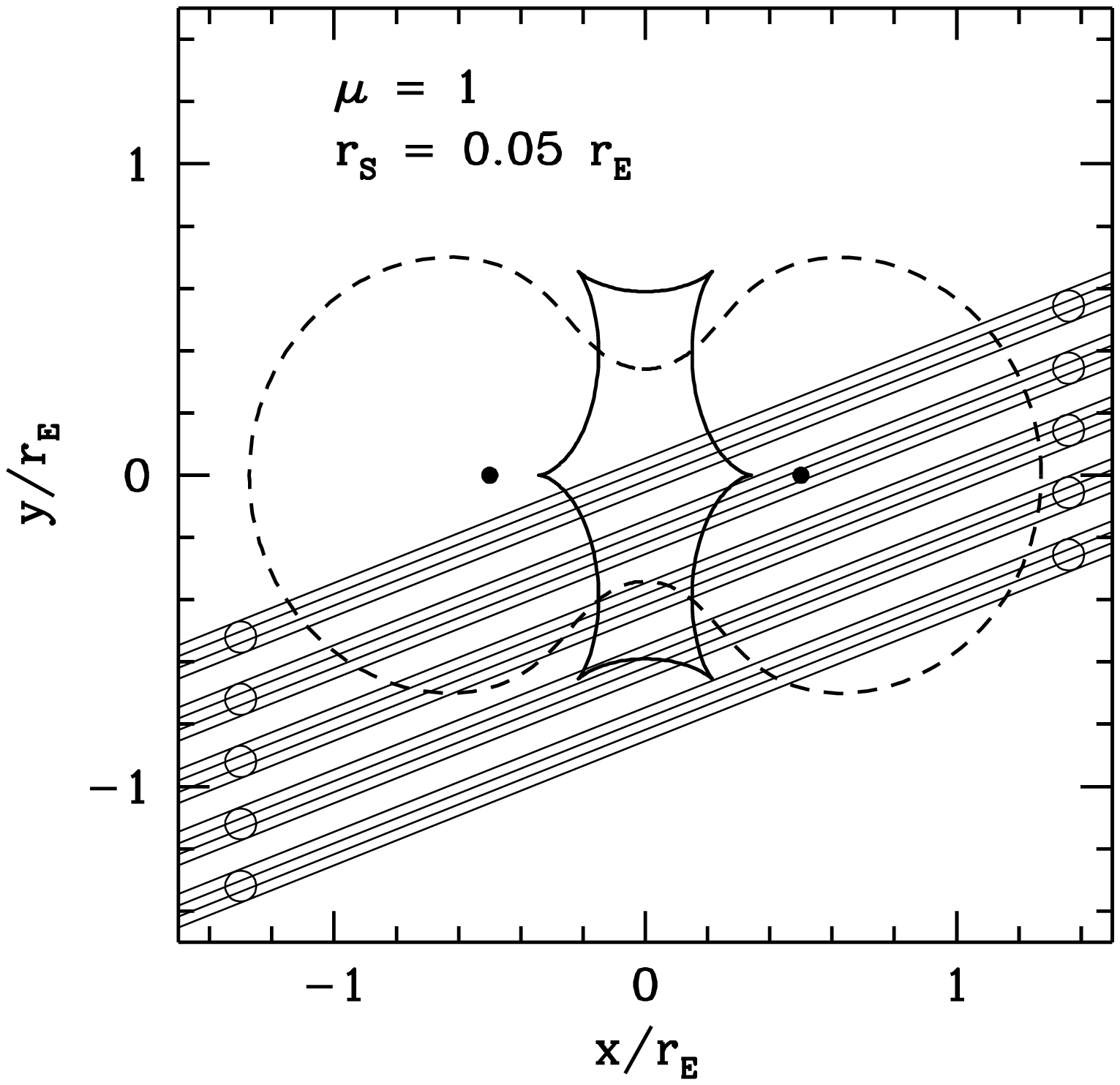}
\caption{\small
The geometry of gravitational microlensing responsible for the light 
curves presented in Fig. 6 are shown.  The two identical point
masses, $ M_1 = M_2 = 0.5 M $, are indicated with two points separated
by one Einstein ring radius $ r_E $.
The closed figure drawn with a thick solid line is the caustic located
in the source plane.
The closed figure drawn with a thick dashed line is the critical curve.
A source placed on a caustic creates an image located on the critical curve.
Five identical sources are moving along the straight trajectories, as 
marked.  All sources have radii
equal $ r_s = 0.05 r_E $, as shown with small open circles.
}
\end{figure}

A few examples of light curves generated by models of
a double lens are shown in Fig. 6, with the geometry of lensing
indicated in Fig. 7.  The model binary is composed of two identical 
point masses (shown in Fig. 7 with two large points):
$ M_1 = M_2 = 0.5 M $, with the separation equal to the Einstein ring
radius corresponding to $ M $, the total binary mass.
The complicated closed figure drawn with a solid line is the caustic.
When a source crosses the caustic then two images appear or disappear
somewhere on the critical curve, shown with a dashed line.
The critical line is defined as the location of all points in
the lens plane which are mapped by the lensing onto the caustics
which is located in the source plane.
When the source is outside the caustic then three images are formed:
one outside of the critical line, and two inside,
usually very close to one of the two point masses.  When the source
is inside the region surrounded by the caustic then the additional two images
are present, one inside and the other outside the region surrounded by
the critical curve.  In our example the five sources, marked with small open
circles, have radii equal to 0.05 of the Einstein ring radius.  The five
straight trajectories are also marked.

The mass ratio of a double lens may be very extreme if one of the
two components is a planet.  Mao \& Paczy\'nski (1991)
proposed that microlensing searches may lead to the discovery of
the first extra--solar planetary system.  That suggestion turned out
to be incorrect as the first extra--solar system with 3 or even 4 planets
has been discovered by other means (Wolszczan \& Frail 1992,
Wolszczan 1994).  Still, it is possible that numerous planetary systems
will be detected with their microlensing effects.  Gould \& Loeb (1992)
and Bolatto \& Falco (1994) made some attempts to refine the probability
of detection as estimated by Mao \& Paczy\'nski (1991), but the real
problem is setting up a practical detection system.  Here we shall consider
just one aspect of the problem, with more discussion to follow in section 7.

Although a search for Jupiterlike objects might prove very interesting,
the detection of Earthlike planets would be far more exciting.  In fact,
such planets have been discovered around the radio pulsar PSR B1257+12
by Wolszczan \& Frail (1992).  Until recently many searches for 
Jupiter-mass planets yielded negative results, leading to the conclusion
that ``... the absence of detections is becoming statistically
significant ... '' (Black 1995).  A recent discovery of a Jupiter mass
companion to a nearby star 51 Pegasi by Mayor and Queloz (1995),
and the discovery of similar companions to the stars 70 Vir and 47 UMa
by Marcy \& Butler (1996) and by Butler \& Marcy (1996), respectively,
demonstrate that such planets exist.

If all stars had Jupiters at a distance of a few astronomical units
then a few percent of all microlensing events might show a measurable
distortion of their light curves (Mao \& Paczy\'nski 1991, Gould \&
Loeb 1992, Bolatto \& Falco 1994).  If a small fraction of all stars
has Jupiters at such distances then the fraction of microlensing
events which may be disturbed by giant planets is correspondingly reduced.
According to Butler \& Marcy (1996)
$ \sim $ 5\% of all stars have super-Jupiter
planets within 5 astronomical units.
Note, that the duration of the likely disturbance is of
the order of $ \sim 1 $ day, i.e. very frequent observations
are necessary in order not to miss them.  

The detection of Earthlike planets
is beyond the range of traditional techniques: radial
velocity as well as astrometric measurements are not accurate
enough, while pulsar timing
is obviously not applicable to ordinary main sequence stars.  It is
interesting to check if Earthlike planets are within reach for
gravitational lensing searches.  
The Earth mass is about $ 3 \times 10^{-6} ~
M_{\odot} $, and a typical star is somewhat less massive than the sun.
Therefore, we consider now a very extreme mass ratio: 
$ M_2/M_1 = 10^{-5} $.

An isolated planetary mass object would create an Einstein ring of its 
own, with the radius $ r_{\rm EP} = r_{\rm E} \times (M_2/M_1)^{1/2} $
However, the same object placed close to a star, develops a complicated
small scale magnification pattern superimposed on the large scale pattern 
generated by the star.  It turns out that this effect is much more 
pronounced than a small disturbance which a planet may cause near the 
peak of the stellar magnification.
An example of the planetary disturbances is shown in Figure 8, in which
a very high magnification stellar microlensing event with the impact
parameter equal to zero is disturbed by eight Earthlike planets placed
along the source trajectory.  Needless 
to say the disturbance caused by each planet is practically independent 
of the disturbances caused by all other planets.

Naturally, this is a very artificial
arrangement, but it makes it possible to present a variety of
microlensing effects of Earthlike planets in a single figure.
The dimensionless time of a planetary disturbance $ t/t_{\rm 0} $
is equal to the dimensionless position of the source $ r_{\rm s}/r_{\rm E} $,
and both are related to the dimensionless location of the planet with
the eq. (8a): $ t/t_{\rm 0} = r_{\rm s}/r_{\rm E} = r_{\rm p}/r_{\rm E}
- r_{\rm E}/r_{\rm p} $.
The planets located close to the star, at $ r_{\rm p} / r_{\rm E} < 1 $,
create local minima in the microlensing light curve presented in
Figure 8; they do this by reducing the
brightness of the image corresponding to $ I_1 $ in Figure 2.
The planets located farther away from the star, at 
$ r_{\rm p} / r_{\rm E} > 1 $ create local maxima (or double maxima)
in the microlensing light curve presented in Figure 8; they do this
by splitting the image corresponding to $ I_2 $ in Figure 2, and 
enhancing the combined brightness.  If a planet is located close the
Einstein ring, i.e. if $ r_{\rm p} \approx r_{\rm E} $ then it affects
the peak of stellar microlensing light curve by disturbing one of the
two images.  While these disturbances are moderately large for Jupiterlike
planets (Mao \& Paczy\'nski 1991) they turn out to be very small for
Earthlike planets.  A detailed description and the
explanation of these phenomena is provided by Bennett \& Rhie (1996).

\begin{figure}[t]
\vspace{7cm}
\includegraphics{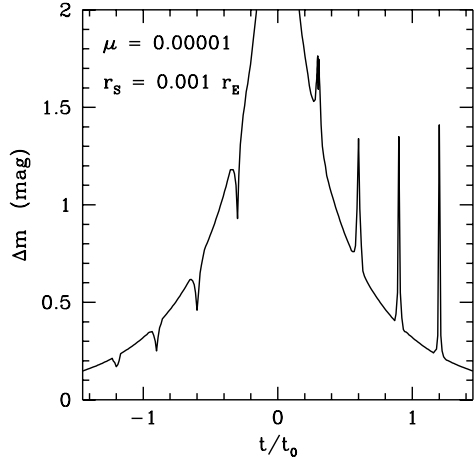}
\caption{\small
Variation of the magnification by a planetary system is
shown as a function of time.  The system is made of a star and
eight planets, each with the mass fraction $ \mu = 10^{-5} $, all
located along a straight line.  The source with a radius 
$ r_{\rm S} = 10^{-3} \, r_{\rm E} $ is moving along the
line defined by the planets, with the impact parameter equal
to zero.  The planets are located at the distances from the star:
$ r_{\rm p } / r_{\rm E} =
0.57, ~ 0.65, ~ 0.74, ~ 0.86, ~ 1.16, ~ 1.34, ~ 1.55, ~ 1.76 $
in the lens plane, which corresponds to the disturbances in
light variations at the times $ t/t_0 = -1.2, ~ -0.9, ~ -0.6,
~ -0.3, ~ 0.3, ~ 0.6, ~ 0.9, ~ 1.2 $, as shown in the Figure.
Note, that planetary disturbances create local light minima for
$ r_{\rm p } / r_{\rm E} < 1 $ ($ t/t_0 < 0 $), and local maxima for
$ r_{\rm p } / r_{\rm E} > 1 $ ($ t/t_0 > 0 $)
}
\end{figure}

\begin{figure}[p]
\vspace{7cm}
\includegraphics{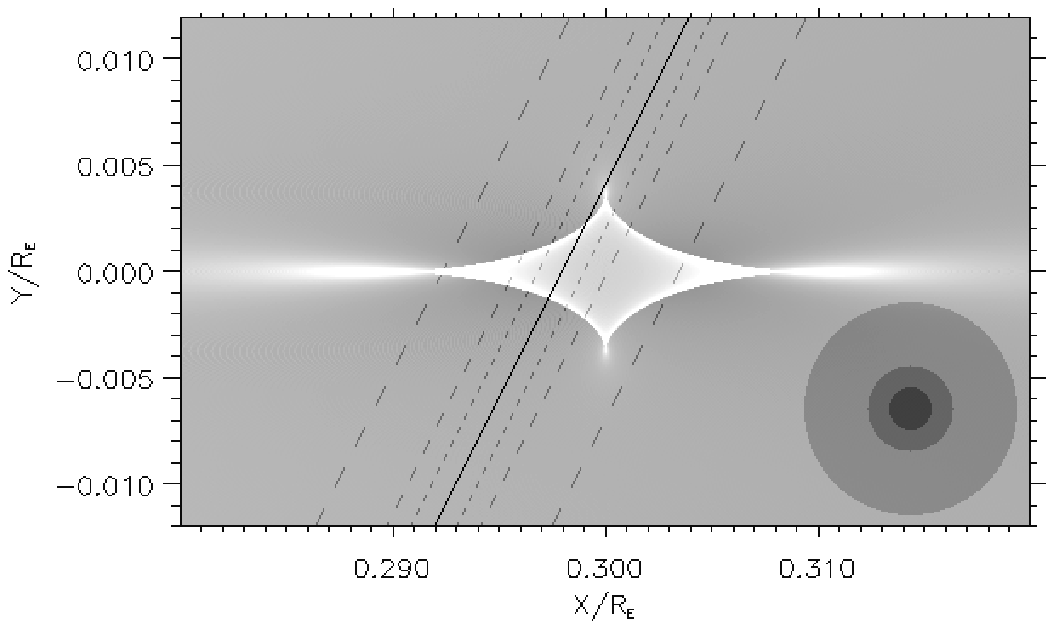}
\caption{\small
The illumination pattern in the source plane created
by a planet with the mass fraction $ \mu = 10^{-5} $.
The planet is located at the distance $ r_{\rm p} / r_{\rm E} = 1.16 $
from the star in the lens plane, which corresponds to 
$ r_{\rm p} / r_{\rm E}= 0.3 $ in the source plane.  
The bright rims are the caustics.  The centers of
three circular sources are moving upwards along the solid straight line,
with the dashed lines indicating the trajectories of the source
edges.  The sources have radii
$ r_{\rm S} / r_{\rm E} = 0.001, ~ 0.002, ~ 0.005 $,
and their sizes are shown in the lower right corner.  
The brightness variations caused by this planetary 
microlensing event are shown in Figure 10.
Note that the light curve shown in Figure 8
corresponds to a source with the radius
$ r_{\rm S} / r_{\rm E} = 0.001 $ moving along the X-axis.
Also note that the area significantly disturbed by the planet
is larger than $ \pi \mu ^{1/2} $, which is the microlensing
cross section for an isolated planet.
}
\vspace{7cm}
\includegraphics{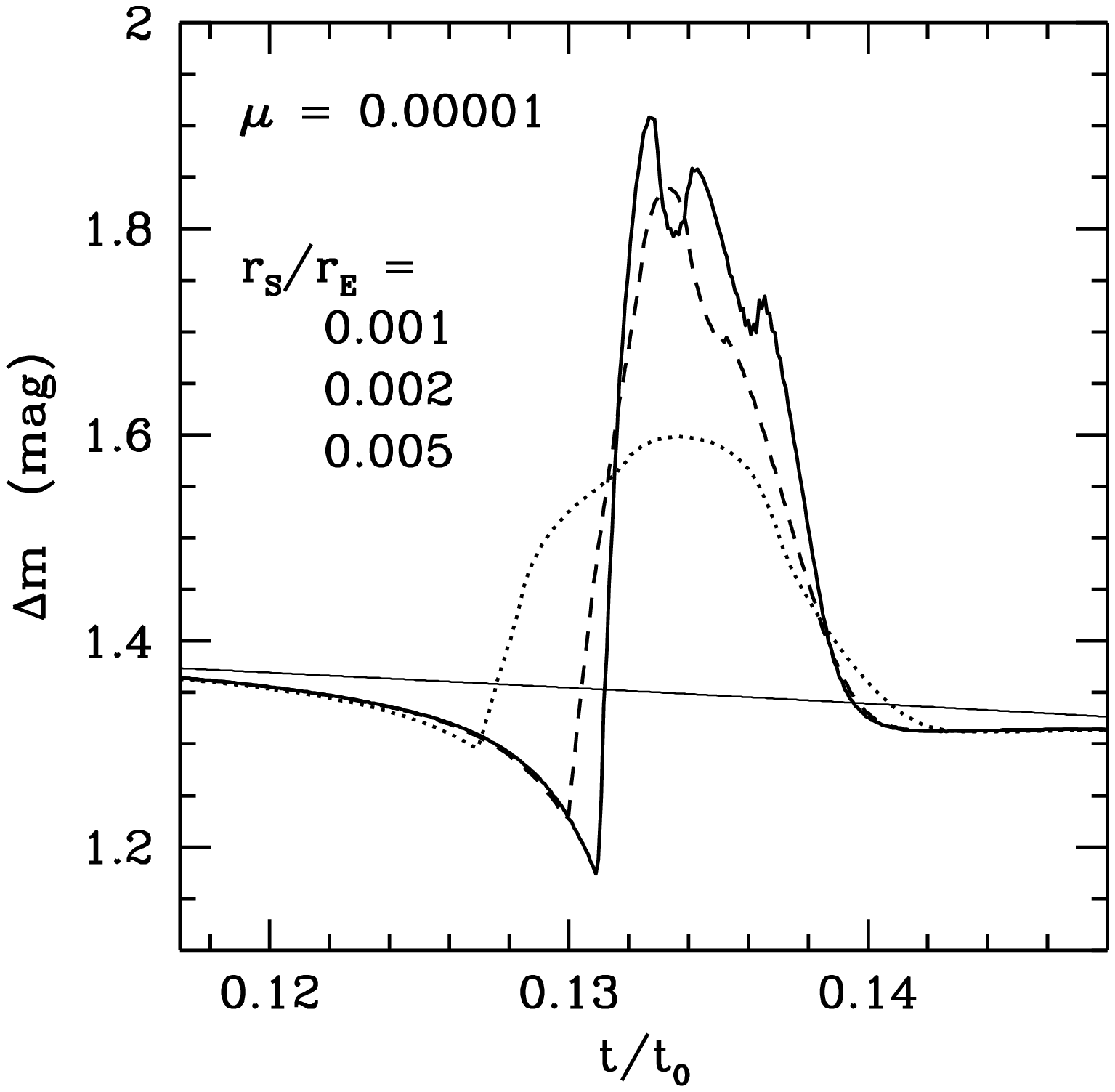}
\caption{\small
Variations of the magnification shown in stellar magnitudes 
as a function of time.
The three light curves, shown with thick solid, dashed, and dotted lines,
are caused by an Earthlike planet with the mass fraction $ \mu = 10^{-5} $,
and they correspond to the three sources with the radii 
$ r_{\rm S} / r_{\rm E} = 0.001, ~ 0.002, ~ 0.005 $, following
trajectories shown in Figure 9.  The thin, slowly descending solid line
corresponds to the fragment of the stellar microlensing light curve
in the absence of planetary disturbance.  The full time interval shown
in this figure corresponds to $ \sim $ 10 hours, and $ t_0 = 0 $
corresponds to the peak of the stellar magnification light curve.
}
\end{figure}

A close-up view of a single planetary event is presented in Figures 9 and 10.
The two-dimensional magnification pattern is shown in Figure 9, together
with the trajectories of three sources.  The corresponding magnification
variations are shown in Figure 10.  This event corresponds to the magnification
disturbance at $ t/t_0 = 0.3 $ in Figure 8, but the source trajectory
is different in the two cases: it was along the X-axis in Figure 8, but it
it is inclined to the X-axis at an angle $ \sim 64^o $ in Figures 9 and 10.
The planetary event lasts longer in Figure 8 than in Figure
10 because the region of high magnification is strongly stretched along
the X-axis, as shown in Figure 9.  This stretching enhances the cross
section for planetary microlensing.  The diversity of
possible light curves is very large, just as it is in the case of 
a binary lens (cf. Figure 6).

Figures 8, 9, and 10 presented in this section were prepared by Dr. Joachim
Wambsganss.  I owe my insight into the diversity of planetary microlensing
phenomena to Dr. David Bennett and Dr. Jordi Miralda-Escud{\`e}.

\section{Various complications}

It is often claimed that gravitational microlensing of stars is achromatic,
it does not repeat, and a symmetric light curve is described by
a single dimensionless quantity: the ratio of the impact parameter
to the Einstein ring radius.  While it is true that the above
description is a good approximation to the majority
of microlensing events, it is well established that none of 
these claims is strictly correct.

Stars are not distributed randomly in the sky, they are well known
to be clustered on many scales.  Most stars are in binary systems,
with the separations ranging from physical contact to 0.1 pc.
Many stars are in multiple systems.  Note that one of the brightest
stars in the sky, Castor, is a sixtuple system.  This applies to
the stars which are sources as well as to those which are lenses.
The case of a double source is simple, as it generates a linear
sum of two single lens light curves (Griest \& Hu 1992).  Naturally, as the
two stellar components may have different luminosities and colors
the composite light curve may well be chromatic.  If the two source
stars are well separated and the lens trajectory is along the line
joining the two, we may have a perception of two microlensing events
separated by a few months or a few years, but both acting on apparently
the same star, as images of the two binary components are unresolved.
This would have the appearance of a recurrent microlensing event, and
it may affect a few percent of all events (Di Stefano \& Mao 1996).

A much more complicated light curve is generated by a double lens as described
in the previous sub--section.  Note, that at least two types of double
lensing events are expected.  In the ''resonant lensing'' case, i.e.
when the two point lenses of similar mass
are separated by approximately one Einstein
ring radius, caustics are formed and dramatic light variations are
expected, as in the case of OGLE \#7 (Udalski et al. 1994d)
and DUO \#2 (Alard et al. 1995b).

As microlensing events are very rare, all current searches are done in
very crowded fields in order to measure as many stars as possible in
a single CCD frame.  This means that the detection limit is set not
by the photon statistics but by overlapping of images of very numerous
faint stars.  A typical seeing disk is about one arcsecond across,
while the cross section for microlensing is about one milliarcsecond
(cf. eq. 13).  Therefore, an apparently single stellar image may typically
be a blend of two or more stellar images which are separated by less
then an arcsecond but by more than a milliarcsecond, i.e. only one
of the two (or more) stars contributing to the blended image is
subject to microlensing.  This implies that when
a theoretical light curve is fitted to the data it should always include
a constant light term (Di Stefano \& Esin 1995).
The contribution from a constant light was found in the double
lenses OGLE \#7 (Udalski et al. 1994d) and DUO \#2 (Alard et al. 1995b),
and with at least one single lens
OGLE \#5 (S. Mao, private communication).  As the various components to
the apparently single stellar image may differ in color, a microlensing
event of only one of them may appear as chromatic (Kamionkowski 1995).
Of course, there should be a linear relation between the variable components
in all color bands.

No star is truly a point source.  The finite extent of a star 
affects a microlensing light curve when the impact parameter of
a single lens is smaller than the source diameter, and also when
the source crosses a caustic of a double lens.  This complicates the
light curve by adding one more adjustable parameter.  This effect,
when measured, may be used to calculate the relative proper motion
of the lens -- source system (Gould 1994a).  It may also be used to study
the distribution of light across the stellar disk, the limb darkening,
and the spots.

So far we assumed that the relative motion between a lens and a source
is a straight line.  In fact the trajectory may be more complicated
as the Earth motion is accelerated by the sun, and in case of a double
lens the two components orbit each other.  This nonlinear motion of
Earth and of binary components makes 
a single lensing event asymmetric if its time scale is longer 
than a few months (Gould 1992).  A complicated double lens
event becomes even more complicated.

\section{Results from current searches}

\subsection{Highlights}

The most outstanding result of the current searches for microlensing
events was the demonstration that they were successful.  By the time this
article is printed over 100 microlensing events will have been 
discovered, mostly by the MACHO collaboration, making this the most
numerous class of gravitational lenses known to astronomers.  Just a 
few years ago, when the projects had been started, the scepticism 
about their success was almost universal, yet at least three teams,
DUO, MACHO, and OGLE have numerous and believable candidate events.
An example of a single microlensing event, OGLE \#2, is shown in 
Fig. 11 following Udalski et al. 1994a).  The microlensing event
took place in 1992.  This object was in the overlap area of two separate
fields, so it had a large number of measurements 

\begin{figure}[p]
\vspace{8cm}
\includegraphics{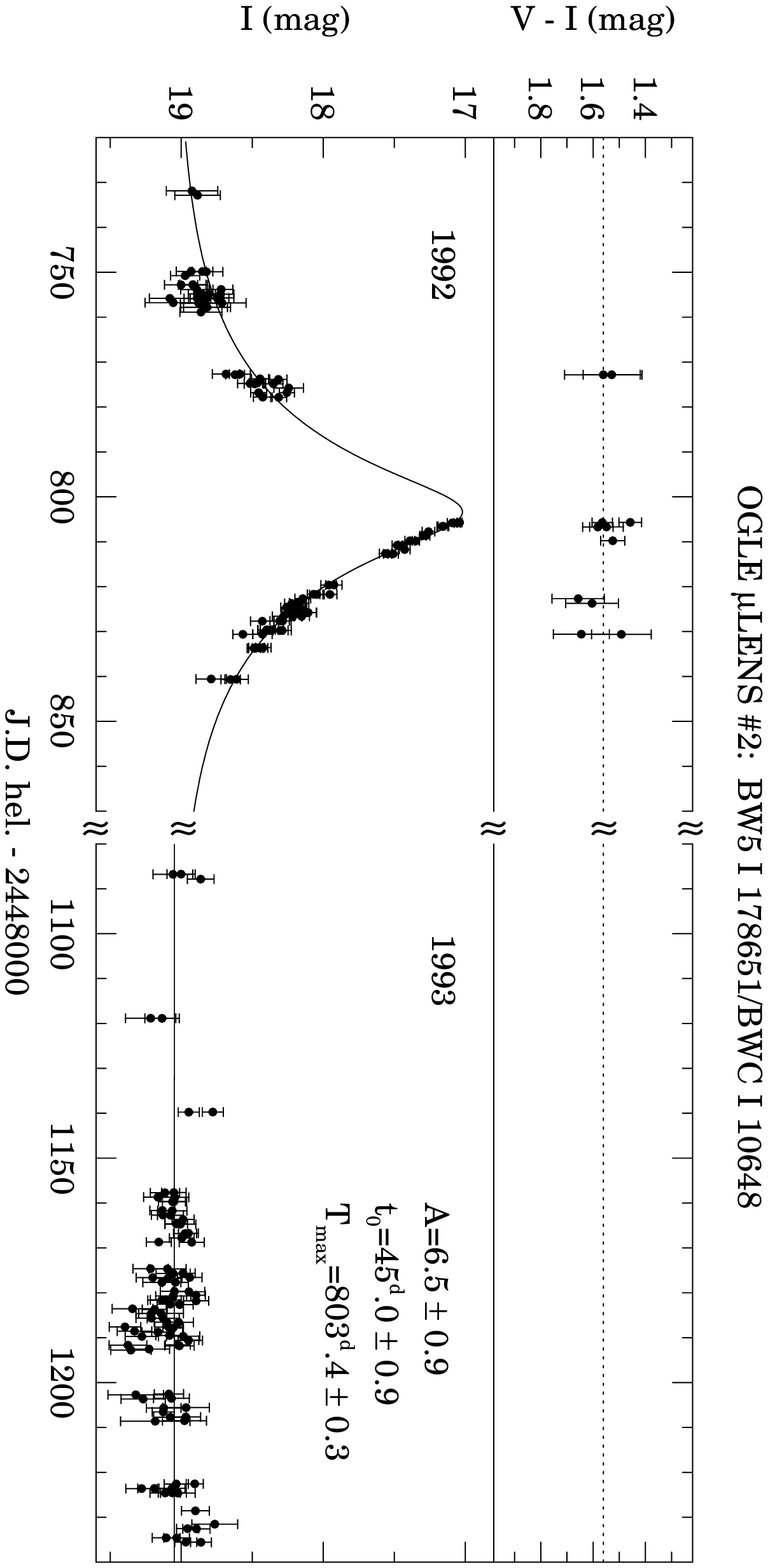}
\caption{\small
An example of the observed light curve due to a single point mass 
lensing: the OGLE lens candidate \#2 (Udalski et al. 1994a).
}
\vspace{8.5cm}
\includegraphics{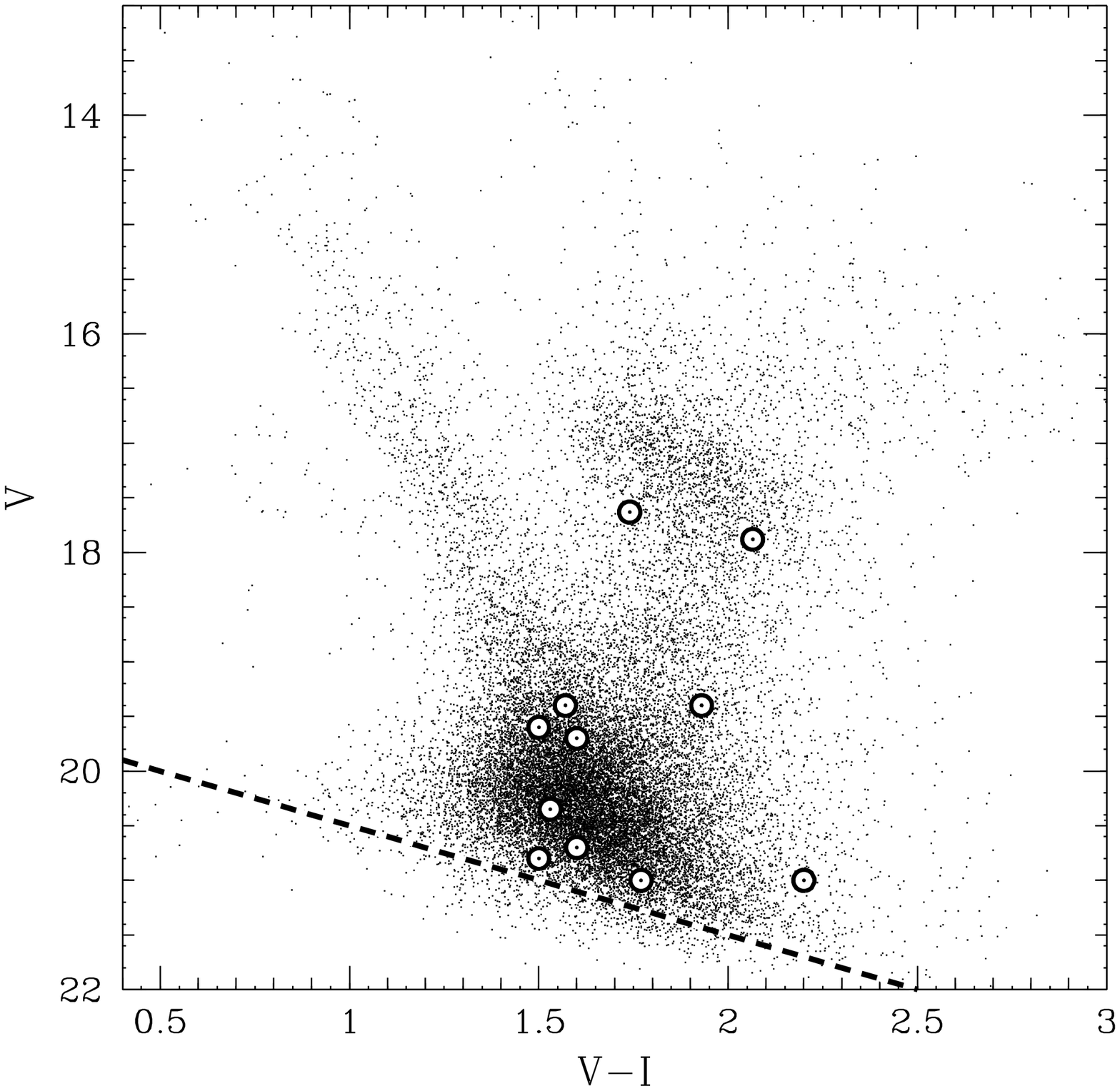}
\caption{\small
The color -- magnitude diagram for stars in Baade's
Window.  The location of 11 OGLE lens candidates is shown with circles.
The dashed line indicates the detection limit applied for
the search (Udalski et al. 1994b).  The majority of stars are at
the main sequence turn -- off point of the galactic bulge,
near $ V-I \approx 1.6 $, $ V \approx 20 $.  The bulge red clump
stars are near $ V-I \approx 1.9 $, $ V \approx 17 $.  The disk
main sequence stars form a distinct band between $ (V-I,V) \approx
(1.5, 19) $ and $ (V-I,V) \approx (1.0, 16) $ (Paczy\'nski et al 1994,
and references therein).
}
\end{figure}

\noindent
in the I--band: 93, 187,
and 94 in the observing seasons 1993, 1994, and 1995, respectively.
The stellar luminosity was constant during these three years, with the 
average I band magnitude of 19.07, 19.10, and 19.13, respectively; the 
standard deviation of individual measurements was 0.13, 0.10, and 0.09 
magnitude, respectively (M. Szyma\'nski, private communication).

The distribution of the OGLE events in the $ (V-I) - V $ color -- magnitude
diagram of the stars seen through Baade's Window, i.e. at
$ ( l,b) \approx (1,-4) $, is shown in Fig. 12.  Note that
the events are scattered over a broad area of the diagram populated
by galactic bulge stars, in proportion to the local number density
of stars multiplied by the local efficiency of event detection
(Udalski et al. 1994b).  A very similar distribution for over 40
MACHO events detected towards the galactic bulge was shown by
Bennett et al. (1995).  It is interesting that three out of $ \sim 40 $
MACHO events appear
to be at the location which is occupied by the bulge horizontal branch
stars as well as the disk main sequence stars, near $ (V-I) \approx 1.2 $,
$ V \approx 17 $.  Spectroscopic analysis should be able to resolve
this ambiguity, i.e. are the lensed stars in the disk or in the bulge.

The most dramatic result is the estimate by the MACHO collaboration
that the optical depth to microlensing through the galactic halo
is only $ 9_{-5}^{+7} \times 10^{-8} $ (based on 3 events), and can 
contribute no more than
20\% of what would be needed to account for all dark matter in the halo
(Alcock et al. 1995a).  The most surprising result is the OGLE discovery
that the optical depth 
is as high as $ 3.3 \pm 1.2 \times 10^{-6} $ (based on 9 events) 
towards the galactic bulge (Udalski et al. 1994b).
This result was found independently by MACHO with 4 events (Alcock et al.
1995c), and qualitatively confirmed by DUO (Alard 1996b).
It should be remembered that all quantitative analyses of the optical 
depth as published to date are based on only $ 16 $ events mentioned
above.  The statistics is improved with the recent preprint
(Alcock et al. 1995e) with the analysis of 45 MACHO events detected
in the direction of the galactic bulge.

Some nonstandard effects which had been first predicted theoretically 
were also discovered.  These include
very dramatic light curves caused by stellar sources crossing caustics
created by double lenses (OGLE \#7: Udalski et al. 1994d, Bennett et al.
1995, DUO \#2: Alard et al. 1995b, and probably one of the MACHO 
Alert events: Pratt et al. 1996), as predicted by Mao \& Paczy\'nski (1991);
the paralactic effect of Earth's orbital motion (Alcock et al. 1995d) as 
predicted by Gould (1992); the light curve distortion of a very high
magnification event by a finite source size (Alcock, private communication)
as predicted by Gould (1994a), Nemiroff \& Wickramasinghe (1994), Sahu (1994),
Witt \& Mao (1994), and Witt (1995); the 
chromaticity of apparent microlensing caused by blending of many stellar 
sources, only one of them lensed (Stubbs 1995, private communication)
as predicted by Griest \& Hu (1992).

The first theoretical papers with the estimates of optical
depth towards the galactic bulge (Griest et al. 1991, Paczy\'nski 1991)
ignored the effect of microlensing by the galactic bulge stars.  That
effect was noticed to be dominant by Kiraga \& Paczy\'nski
(1994), who still ignored the fact that there is a bar in the inner
region of our galaxy (de Vaucouleurs 1964, Blitz \& Spergel 1991).
Finally, the OGLE results forced upon us the reality of the
bar (Udalski et al. 1994b, Stanek et al. 1994, Paczy\'nski et al. 1994, 
Zhao et al. 1995, 1996).  This ``re-discovery'' of the galactic bar by
the microlensing searchers, who were effectively ignorant of its existence, 
demonstrates that the microlensing searches are becoming a useful new 
tool for studies of the galactic structure (Dwek et al. 1995).
We are witnessing a healthy interplay between 
theory and observations in this very young branch of astrophysics.

An example of a dramatic light curve of the first double
microlensing event, the OGLE \#7, is shown in Fig. 13 for 1992 and 1993
following Udalski et al. (1994d).
This star was found to be constant in 1992, 1994, and 1995.  The 
average magnitude based on 32, 45, and 41 I-band measurements in these three
observing seasons was 17.53, 17.52, and 17.54, respectively, with
the variance of single measurements being 0.07, 0.04, and 0.03 magnitudes,
respectively (M. Szyma\'nski, private communication).  The objects was also
found in the MACHO database (Bennett et al. 1995), confirming the
presence of the second caustic crossing event near JD 2449200, and
demonstrating that the light variation was achromatic.

\begin{figure}[t]
\vspace{8cm}
\includegraphics{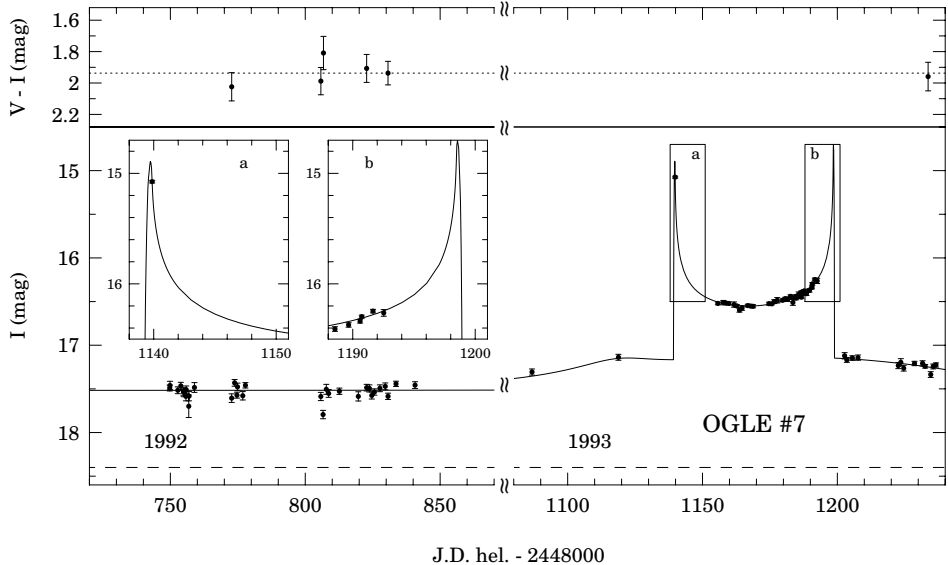}
\caption{\small
An example of a binary lensing: the OGLE lens candidate \#7
(Udalski et al. 1994d).  The region of the two caustic crossings, (a)
and (b), is shown enlarged in the two inserts.  The MACHO collaboration
has a few dozen additional data points in two bands demonstrating that
the light variations were achromatic; three MACHO data points cover
the second caustic crossing (b) (Bennett et al. 1995).
}
\end{figure}

It is hard to decide which was the first microlensing event, as this
depends on what ``the first'' is supposed to mean.  If we take the
time at which a microlensing event reached its maximum, then the first
was OGLE \#10 which peaked on June 29, 1992 (Udalski et al. 1994b).  It was
followed by six other OGLE events that were observed that summer.  However,
these were not uncovered until the spring of 1994, when
the automated computer searches finally caught-up with the
backlog of unprocessed data.
The OGLE collaboration discovered its first
event, OGLE \#1 on September 22, 1993, but it peaked on June 15, 1993,
almost a full year later than OGLE \#10.
The first event to be ever noticed by a human was MACHO \#1 --
Will Sutherland saw it come out of a computer on Sunday, September 12,
1993 (C. Alcock, private communication).
The first three papers officially announcing the first computer
detections were
published almost simultaneously by EROS (Aubourg et al. 1993), by
MACHO (Alcock et al. 1993), and by OGLE (Udalski et al. 1993).  

Unfortunately, it is likely that the two events reported by EROS might 
have been due to intrinsic stellar variability rather than microlensing.
EROS \#1 has been recently found to be an emission
line Be type star (Beaulieu et al. 1995).
The MACHO collaboration has identified a new class of variable stars,
referred to as bumpers (Cook et al. 1995).
These are Be type stars, and
it is possible that EROS \#1 exhibited a bumper phenomenon.  
EROS \#2 has been recently found to be an eclipsing binary, possibly with 
an accretion disk (Ansari et al. 1995a).  Stars
with accretion disks are known to exhibit diverse light variability,
and the EROS \#2 event might have been due to disk activity.
In any case, the fact that both EROS events were related
to rare types of stars makes them highly suspect as candidates
for gravitational microlensing.  

A total of about 100 gravitational microlensing events has been reported
so far by the three collaborations: DUO, MACHO, and OGLE.  The DUO 
collaboration reported the detection of 13 microlensing events towards 
the galactic bulge, one of them double (Alard 1996b, Alard et al. 1995a,b).
The MACHO collaboration reported at various conferences a total of
about 8 events towards the
LMC and about 60 events towards the galactic bulge, and has many more
in the data that is still analyzed.  The OGLE collaboration has
detected 18 events towards the galactic bulge, one of them double.  In
addition various collaborations confirmed events detected by the others.
I am not aware of a single case in which there would be a discrepancy between
the data obtained by various collaborations, though there is plenty of
difference of opinion about some aspects of the interpretation.
Note, that the quantitative analysis of all those
$ \sim 100 $ events lags behind their discovery.  Nevertheless, we may 
expect that robust determinations of the optical depth towards the galactic
bulge and towards the LMC will be published in the very near future, as the 
data is already at hand.

A major new development in the microlensing searches is the on line 
data processing.
The OGLE collaboration implemented its ``Early Warning System'' (EWS), a
full on-line data processing system 
from the beginning of its third observing season, i.e. from April 1994
(Udalski et al. 1994c).  As a result all 8 OGLE events 
detected in 1994 and 1995 were announced over Internet in real time.
The MACHO collaboration, with its vastly higher data rate, implemented
the ``Alert System'' with partial on-line data processing in the
summer of 1994 (Stubbs et al. 1995), and full on-line processing by the 
beginning of 1995 (Pratt et al. 1996).
The total number of events detected with the MACHO Alert
System as of October 15, 1995 is about 40, and all of them were announced
over Internet in real time.

The first papers based on the follow--up of the real time announcements
of microlensing events have already been published: 
Szyma\'nski et al. (1994) presented the
first light curve of the first event announced by the MACHO Alert System.
Benetti et al. (1995) presented the first spectra for
the on--going microlensing event.

The implementation of PLANET -- Probing Lensing Anomalies NETwork (Albrow
et al. 1996) was a major development in 1995.  The aim of the project is
to follow the announcements of real time detection of microlensing
events (currently implemented by the OGLE and MACHO collaborations)
with frequent multi-color observations on four telescopes: the Perth
Observatory 0.6 m telescope at Bickley, Australia, the 1 m telescope
near Hobart in Tasmania, the South African
Astronomical Observatory 1.0 m at Sutherland, South Africa, and the
Dutch-ESO 0.92 m at La Silla, Chile.

\subsection{Data analysis}

In order to translate the observed rate of microlensing events into
quantitative information about the optical depth and the lens masses it
is necessary to calibrate the detection system.  This is fairly
straightforward, at least in principle, as all data processing is
done with computers, with automated software.  Naturally, for the
calibration to be possible every step of the detection process has
to be done according to well defined rules, following the same algorithm
for the duration of experiment.  The calibration is done by introducing
artificial microlensing events into the data stream and using the same
algorithm to ``detect'' them as the one used for real detections.
This process can be done at two very different levels: pixels or star
catalogs.

If the original data is collected with a CCD camera then from the beginning
it is stored in a computer memory in a digital form.
Every CCD image obtained by OGLE had $ 2048 \times 2048 $ pixels for
a total of 8 Megabytes of pixel data.  Every
MACHO exposure generated 8 such CCD frames, 4 in each of two color bands,
for a total of 64 Megabytes.  On a clear night between 30 and 100 exposures
may be taken.  The OGLE collaboration was given $ \sim 70 $ nights on
the Swope 1 meter telescope at the Las Campanas Observatory in Chile
(operated by Carnegie Institution of Washington) for each of
the four observing season: 1992--95, and generated $ \sim 20 $ Gb of data
every year.  The MACHO collaboration refurbished a dedicated 1.3 meter 
telescope at the Mount Stromlo Observatory in Australia.  Their observations
began in mid-1992, but the routine operation started in January 1993.
Currently, the MACHO collaboration obtains $ \sim 800 $ Gb of data every 
year.  

If the original
data is obtained with photographic plates, as it was done for
the main part of the EROS project and the whole DUO project, then 
the plates have to be scanned and digitized before the data
can be stored in a computer memory.  A single Schmidt plate
has $ 28,000 \times 28,000 $ pixels for a total of $ \sim 1.6 $ Gb of data.

Once the data is available in a form of digital pixel images
a dedicated software is used to measure the location and
the brightness of stellar images.  The MACHO and OGLE collaborations
modified DoPhot (Schechter et al. 1993) to make it faster (Udalski et 
al. 1992, Bennett et al. 1993).  The coordinates of stellar 
images were determined on good CCD frames which were chosen to be the
templates.  All other frames were first shifted to coincide with 
the templates using a few bright stars, and the coordinates of all
other stars were adopted from the template.  Next, the brightness of
all template stars was measured on all frames.  This speeded the
data processing by a large factor, but restricted photometry to those
stars which were found on the templates.  Also, this procedure
makes it impossible to detect proper motions.  Using a modified DoPhot
software up to $ 2 \times 10^5 $ stars could be measured on
a CCD frame with $ \sim 4 \times
10^6 $ pixels, which corresponds to the effective number of $ \sim 20 $
pixels per star.  The DUO software (Alard 1996b, Alard et al. 1995a)
measured not only the brightness
but also the position of every stellar image on every digitized Schmidt
plate, for a total of $ \sim 1.4 \times 10^7 $ stars per plate, i.e.
$ \sim 56 $ pixels per star.

The results of stellar photometry are combined into a database of
photometric measurements which is searched for microlensing events.
Very stringent criteria have to be used for detection because the 
events are very rare.  For example, one of the OGLE 
conditions was the requirements that at least 5 consecutive measurements
of the candidate object had to be brighter than normal, well measured
brightness of the star, by more than three standard deviations
(Udalski et al. 1994b).  This conservative approach was necessary
in order not to be swamped with fictitious ``events'', but it made the
efficiency of the detection rather low.  Let $ N $ be the total number of 
photometric measurements of all stars, $ n $ the number
of microlensing events detected in the database, and $ \tau $ be the
optical depth to microlensing as estimated on the basis of these detections.
Using the published MACHO and OGLE results one finds that $ N \tau / n \approx
50 - 100 $.  This implies that the effective
number of photometric measurement needed for a detection of
a single event was $ 50 - 100 $.

The first catalog level estimate of the efficiency of microlensing
detections was published by the OGLE collaboration (Udalski et al. 1994b).
It was done introducing artificial microlensing events into the
database of photometric measurements and following exactly the
same criteria for detection as the criteria used in the original search.
It turned out that the efficiency was approximately constant for events
in the time scale range 30 days $ < t_0 < $ 100 days, but it
dropped rapidly towards shorter time scales: down by a factor 10
at $ t_0 \approx 3 $ days and a factor 100 at $ t_0 \approx 1 $ day.
Very similar results were found by the MACHO collaboration with a much
more thorough ``pixel level'' calibration (Alcock et al. 1995a).  In that 
case the artificial events are added as artificial stars to the images
obtained with the CCD detector.  This means that the effect of
blending of stellar images is automatically taken into account.  The
pixel level calibration is the correct way to proceed, but it is
much more time consuming than the catalog level calibration.  Fortunately,
the results are the same to within $ \sim 20 \% $
as the two opposite effects nearly cancel out: stellar blending means
that there are really more stars that may be subject to microlensing,
and this increases the number of events per stellar image.
On the other hand the apparent brightening is reduced making some
events impossible to detect.

Note, that as long as the procedure used for the detection of artificial
events is exactly the same as the procedure used for the original search,
the estimate of the event rate and the optical depth is not influenced
in a systematic way by the specific criteria.  Naturally, if the criteria
are too stringent then there are fewer detections (in the real data as
well as in the simulations), and the random errors of the estimated values
increase.  If the criteria are too lax then in addition to genuine
microlensing events a number of intrinsically variable stars, or even
artifacts of the detector system, may enter the sample, introducing
uncontrollable systematic errors.  As far as I know we have no
quantitative procedure to define the optimum criteria at this time.

The fact that the current detection systems are sensitive to events over
limited range of time scales implies that the estimates of the optical
depth and/or the event rates can only provide the lower limits.  As the
duration of searches increases so does their sensitivity to ever longer
events.  It will take a different observing procedure
to improve the sensitivity to very short events.  Most searches done
so far made only 1 or 2 photometric measurements per star per clear
night.  The only CCD search with up to 46 photometric measurements per 
star per clear night was done by the EROS collaboration, and the null
result was reported by Aubourg et al. (1995).

\subsection{Consensus and no consensus}

There is a consensus now about a number of important issues
related to microlensing in our galaxy and near our galaxy.  The
most important is the consensus that the microlensing phenomenon has been 
detected.  It is not possible to tell at this time what fraction of
so called ``candidate events'' is real, and what fraction is due
to poorly known types of stellar variability.  My personal guess
is that the contamination fraction is probably below 10\%, or so.
My arguments in favor of microlensing as the dominant source
of the newly detected variability are the following:

\begin{enumerate}

\item 
The observed light curves are achromatic, their shapes are well
described by simple theoretical formulae.

\item
The distribution of magnification factors is consistent with the theoretical
expectations (Udalski et al. 1994a), with some events magnified by a factor 
up to $ \sim 100 $ (Stubbs 1995, private communication).

\item
The double lensing events have been detected, as expected, and roughly
at the expected rate.

\item
The parallax effect has been detected, as expected.

\item 
The spectrum of the only event monitored spectroscopically has been
found to be constant throughout the intensity variation (Benetti et al.
1995).

\item
The galactic bar has been ``rediscovered'' through the enhanced optical depth.

\end{enumerate}

There is also a consensus on some science issues:

\begin{enumerate}

\item
The optical depth towards the galactic bulge is large, 
$ \sim 3 \times 10^{-6} $.

\item
The optical depth towards the LMC is small, $ \sim 10^{-7} $.

\item
A lot of very interesting science not related to microlensing
can be done with huge databases generated with the searches.
These include color magnitude diagrams as well as many types
of variable stars.

\end{enumerate}

On some issues there is no consensus yet:

\begin{enumerate}

\item
What is the location of objects which dominate
lensing observed towards the galactic bulge?  Are these
predominantly the galactic bulge stars (Kiraga \& Paczy\'nski 1994,
Paczy\'nski et al. 1994, Zhao et al. 1995, 1996), or are the lenses mostly
in the galactic disk (Alcock et al. 1995c)?

\item
What is the dominant location of the objects responsible for the
lensing observed towards the LMC?  Are these in the galactic disk,
galactic halo, the LMC halo, or in LMC itself?  Are they stellar
mass objects or are they sub-stellar brown dwarfs (Sahu 1994,
Alcock et al. 1995a)?

\item
What fraction of microlensing events is caused by double lenses?

\end{enumerate}

A consensus on these issues is likely to emerge within a few years,
but no doubt new controversies will develop as the volume of data increases.

\subsection{Serendipity results}

The microlensing searches worked because practical implementation
of massive data acquisition and processing systems became possible.  These
searches generated huge databases of multi band photometry of millions
of stars, and lead to
the discovery of thousands of variable stars, most of them
new.  The OGLE galactic bulge variable star catalogues published so far
contain full data for 1656 pulsating, eclipsing and other short period
variables, including their coordinates and finding charts
(Udalski et al. 1994e, 1995a,b).  The other OGLE papers include data
on RR Lyrae type stars in the Sagittarius dwarf galaxy (Mateo et al 1995b) 
and Sculptor dwarf galaxies (Ka\l u\.zny et al. 1995a).
The MACHO collaboration presented period -- luminosity diagram for
$ \sim 1500 $ cepheids in the LMC and identified 45 double mode
pulsators among them (Alcock et al. 1995b), and has a total of
$ \sim 90,000 $ variables in its archive (Cook et al. 1995).
The EROS collaboration published data on 80 eclipsing binaries
in the LMC (Grison et al. 1995, Ansari et al. 1995a).  The DUO
collaboration has data on $ \sim 15,000 $ galactic bulge
variables (Alard 1996a).  These include $ \sim 1,200 $ pulsating stars
of the RR Lyrae type ab in the galactic bulge, and $ \sim 300 $
such stars in the Sagittarius dwarf, considerably extending
the known size of that galaxy, and clearly demonstrating the usefulness
of variable stars as tracers.

Perhaps the most important serendipity result is the discovery
by Ka\l u\.zny et al. (1995b) of the first detached eclipsing
binaries at the main sequence turn--off point of the globular cluster
Omega Centauri.  The follow--up spectroscopic observations will determine,
for the first time, the masses of stars at the globular cluster
main sequence turn--off point, and hence will provide a sound basis
for the reliable age and helium content determinations.  The point is
that while the theoretical color--magnitude diagrams are affected
by our lack of understanding of the mixing length theory, the 
mass--luminosity relation is insensitive to large changes of the
mixing length (Paczy\'nski 1984).

The same detached binaries will allow very accurate determination
of distances to globular clusters, while the follow--up spectroscopic
observations of the detached eclipsing binaries discovered by
the EROS collaboration (Grison et al. 1995) will provide a very accurate
distance to the LMC (Paczy\'nski 1996).  We may expect that in the near 
future detached eclipsing binaries will provide accurate distances to
all galaxies of the Local Group (Hilditch 1995).
It should be pointed out that in order to discover a detached eclipsing
binary one needs a few hundred photometric measurements.
Only one out of a few thousand stars is of this type, with
deep and narrow primary and secondary eclipses, and no anomalies in the
light curve.  This implies one needs $ \sim 10^6 $ photometric measurements
to detect one good system.  This is a lot of work, but it
will lead to the determination of accurate ages of globular clusters,
their helium content, and the Hubble constant, some of the
most important numbers for cosmology.

Color-magnitude diagrams obtained in the standard (V,I) photometric system
by the OGLE collaboration for the galactic bulge region show evidence
for the galactic bar (Stanek et al. 1994).  They also indicate that the 
galactic disk has a low number density of luminous (i.e. massive and young) 
as well as faint (i.e. low mass and old) main sequence stars in the inner
$ \sim 4 $ kpc (Paczy\'nski et al. 1994), which seems to be consistent
with the presence of the strong bar.  The color--magnitude diagrams
obtained for the recently discovered Sagittarius dwarf allowed Mateo et al.
(1995a) to determine the distance of $ 25.2 \pm 2.8 $ kpc to this galaxy,
and to estimate its age and metallicity to be 10 Gy and [Fe/H] 
$ = -1.1 \pm 0.3 $, respectively.  Alard (1996a) and Mateo et al. (1996)
found a large extension of this galaxy by discovering in it a large number
of RR Lyrae variables, a by product of massive photometry carried out
by DUO and OGLE.

\pagebreak
\section{The future of microlensing searches}

\subsection{The near future}

The near future of the microlensing searches is easy to predict:
more of the same, or rather very much more of the same.  While
the MACHO collaboration has streamlined its data processing and
does it in real time, the EROS and OGLE collaborations are building
in Chile their 1-meter class telescopes to be dedicated to massive
microlensing searches.  Many other groups are either planning or
developing new detection systems.  In particular, a few groups
intend to monitor M31 (Crotts 1992, Ansari et al. 1995b = AGAPE =
Andromeda Galaxy and Amplified Pixels Experiment).  

It is not known how many stars can be monitored from the ground in the
direction of the galactic bulge and all members of the Local Group
galaxies, but a fair estimate is $ \sim 2 \times 10^8 $, i.e. about
one order of magnitude more than the current $ \sim 2 \times 10^7 $ of the
MACHO collaboration.  Note, that stars towards the galactic bulge and 
towards the LMC and SMC are bright, and even with a 1-meter class
telescope the detection limit is set by the crowding of stellar images,
not by the sky or the photon statistics.  Therefore, the better the
seeing the more stars can be detected and accurately measured.
It is easily imaginable that within a few years the detection rate
will increase to $ \sim 300 $ events per year towards the galactic
bulge and $ \sim 10 $ events per year towards the LMC and SMC.
With $ \sim 1000 $ events detected towards the bulge, and $ \sim 30 $
towards the LMC and SMC it will be easy to map the distribution.

The observed distribution of lensing events in the
sky will soon reveal the space distribution of the lenses
towards the galactic bulge (cf. Evans 1994, Kiraga 1994).
A rapid increase of the optical depth towards
the center will indicate that the lenses
are located predominantly in the bulge of our galaxy.
A more or less uniform optical depth will point to the 
galactic disk as the dominant site of the lensing objects.
The task will be more difficult in the case of LMC.  The observed
rate is low, and it will take longer to accumulate good statistics.
A rapid increase of the optical depth towards
the center will indicate that the lenses
are located predominantly in the bulge of LMC.
A more or less uniform optical depth would be more difficult to interpret,
as such a result would be compatible with
the lenses located either in the galactic disk or in the galactic halo.
It may be necessary to measure the optical depth in as many directions 
as possible, towards all members of the
Local Group of galaxies.  This is a difficult task as these are
far away, and hence their stars are very faint.  It is
likely that 2-4 meter class telescopes will be required
for the observations (Crotts 1992, Colley 1995, Ansari et al. 1995b).

It is interesting to note that the statistics of microlensing
of the galactic bulge red clump stars will allow the determination
of the geometrical depth of the bulge/bar system (Stanek 1995).
The lensed objects will be found predominantly among the stars
located on the far side.  Therefore, these stars will appear to be 
fainter, at least on average, than typical red clump stars.

It is likely that within a few years the range of event time
scales between 1 hour and 3 years will be very well sampled.
With the observed distribution of event time scale known over this
whole range, or with stringent upper limits available for some
part of the range, it will become meaningful to translate this
distribution into the mass function of the lensing events.  The
necessary pre-requisite will be the knowledge of the space distribution
and kinematics of the lenses.  It seems safe to expect that the
space distribution will be revealed through the variation of the
optical depth over the sky, while the kinematics will follow from
the improved understanding of the galactic structure.  This analysis
will not be good enough to know the mass of any particular lens to
better than a factor 2 or 3, but it is likely that the overall
range of masses will be known with a reasonable accuracy.  

There are very many papers and preprints written about the possible
distribution of mass within the dark halo, or within the galactic
disk, as well as on the possible nature of the lensing objects
(those include gas clouds and axion miniclusters).  These will not
be reviewed here, as I consider them too speculative.  A more
direct analysis of the data will provide the answers to most
interesting current questions in a matter of a few years.

It is virtually certain that the expansion of microlensing searches
will be followed with the expansion of the follow-up observations.  There
is a very natural division of labor in this area as well as in the
area of supernovae searches.
In order to discover very rare events one needs detectors with as many 
pixels as possible.  The multi--band coverage is not essential.
The pixels are expensive and difficult to purchase, and it is practically
not possible to fill the
whole field of view of a modern 1-meter class telescope with CCDs,
as the field may be 1.5 -- 5 degrees across.  Given a limited number of
pixels we have to decide what is more efficient: to put them all in
one plane, making the search area as large as affordable, or to
cover only 50\% of the area in two bands?  Note that every detection
system is most sensitive in a particular band.  For example the
OGLE system can record more stellar images in I-band than in V--band 
in a fixed exposure time.  Also, when the moon is bright the sky intensity
in the I band is smaller than it is in the V band.  Thin CCD
chips are currently so expensive that extending the search to the
B and U bands is not possible within a realistic budget.  The implementation
of real--time data analysis, coupled to the distribution of
information over the Internet, and the availability of follow-up systems
like PLANET (Albrow et al. 1996) makes the multi-band search unnecessary.

On the other hand it is most essential for the follow-up observations
to be as multi-band and as frequent as possible.  As only a relatively
small number of pixels is needed for the follow-up it should be possible 
to use thin CCD chips for this purpose.  A frequent time coverage is
most essential as many interesting phenomena happen on the time scale on
which a star moves across its own diameter, which takes a few hours.
Such source resolving events may allow the measurement of the relative 
proper motion
in the lens--source system (Gould 1994a), as well as to study the structure
of stellar photospheres.  They may also lead to the detection of planets
around stars (Mao \& Paczy\'nski 1991).

A dramatic improvement in the data processing software may be expected 
when the current photometry of stellar images is supplemented
with a search for variable point sources using digital image subtraction,
also known as ``CCD frame subtraction''.
The image subtraction is very elegant, and its principles are very
simple, as described and implemented by Ciardullo et al. (1990).
Imagine that we have two images of the same area in the sky taken
under identical seeing conditions, and the same atmospheric extinction.
Every point source has an image spread over many pixels according to the
point spread function, PSF.  When the two images are subtracted from
each other there should be no residuals if nothing has changed.  If some
stars changed their brightness then there should be some residuals,
positive or negative, with the profile defined by the PSF.  These
residuals can be measured, and so the stellar variability can be detected.
If we do not care about stars that remain constant then this is by far the
most efficient way to proceed.  Also, this method may allow the detection
of variables which are too crowded to measure the non--variable component
of their light due to the constant contaminating stars.

Unfortunately, it is difficult to implement the image subtraction
technique, as there are many practical problems as described
by Ciardullo et al. (1990).  Therefore, it seems reasonable to
expect that the proof of its practical usefulness should be provided
by the determination of light curves of periodical
variables, like pulsating or eclipsing stars.
It is much easier to detect a periodic signal, and to 
confirm its reality, than to detect convincingly a non--periodic and
non repeating signal, like a microlensing event. 
However, the recently published efforts (Crotts 1992, Baillon et al. 1993,
Ansari et al. 1995b, Gould, 1995b, 1996) seem to concentrate on the
detection of non--periodic signals, referred to as ''pixel lensing''.

\subsection{A more distant future}

Any discussion of a distant future is obviously speculative, but it
is also entertaining.  A rather obvious idea is to observe microlensing
with space instruments.  The natural reasons include stable weather, 
perfect seeing, and access to the UV and IR.  However, there are special
advantages for microlensing in putting an instrument at a distance
of $ \sim 1 $ AU from the ground based telescopes, as this would allow
to detect the microlensing parallax effect as pointed out by Refsdal
(1966) and by Gould (1994b,
1995a).  Any observations done at a single site provide only one quantity 
which has physical significance: the time scale $ t_0 $.  Unfortunately, this
time scale is a function of the lens mass, its distance, and its 
transverse velocity (cf. eq. 15), none of which is known.  However,
the illumination pattern created by the lens varies substantially on
a scale of a fraction of the Einstein ring radius, or its projection
onto the observer's plane.  Hence, even a small space telescope placed at
a solar orbit would reveal a different light curve of the lens, and provide
this way some additional information about the lens properties.  Unfortunately,
this will not be sufficient to solve for all three unknowns.  However,
if the same event happens to resolve the source then the relative
proper motion of the lens--source system can be determined (Gould 1994a),
and perhaps a complete solution could be obtained.  In particular, the
determination of the lens mass might be possible.

There are some problems with this idea.  One of them is the high
cost, likely to be of the order of $ \sim \$10^8 $, as compared
with $ \sim \$10^6 $ for a typical ground based microlensing search
(though supposedly ``dollars in space weigh less'').  Also, even
though the parallax effect alone can provide some constraint on the
lens, it cannot provide a unique determination of the lens mass, with a
possible exception of some special cases,
like the very high magnification
events during which the source is resolved.
Unfortunately, such events are very rare.
Perhaps the single most dramatic impact of the space measurements would be
a direct proof that a particular event is indeed caused by gravitational
microlensing as no other phenomenon can be responsible for the difference
in the observed light curves.

Many of the problems, including a decent statistical determination of
the lens masses, will be solved in the near future 
with the ground based observations.
When the optical depth to the galactic bulge is mapped with $ \sim 10^3 $
lensing events then the space distribution of the lenses (galactic disk
versus galactic bulge/bar) will be definitely understood.  As the kinematics
of the disk and the bulge/bar stars can be directly observed an adequate
model for the lens statistics will be developed as a refinement of the
approach outlined in the section 3.  Any particular lens is not likely
to have its mass determined to better than a factor $ \sim 3 $ or so,
but the average mass, and perhaps the mass range and the slope of the mass
function will be deduced.   This task will be relatively easy for the
lenses observed towards the galactic bulge because of their high rate.
It will take much more time to obtain equally reliable information
about the lenses observed at high galactic latitudes because of their
very low rate.  Yet, continuous
monitoring of the stars in all Local Group galaxies
will provide the basis for the determination of the lens distribution
and masses in a matter of a decade or so.  This is basically the matter
of detecting enough events to map the variation of the
optical depth over the sky.

There are at least two cases in which the masses of individual lenses
can be measured with no ambiguity.  These are lenses belonging to
globular clusters seen against the rich stellar background of the galactic
bulge or LMC/SMC (Paczy\'nski 1994), and the nearby high proper motion
stars seen against the distant stars of the Milky Way or LMC/SMC (Paczy\'nski
1995a).  In both cases the distances and proper motions of the lenses
can be measured directly, or indirectly (the lenses in the globular
clusters will be too dim to see them), and the lens mass remains
the only unknown quantity on the right hand side of the eq. (15).

\pagebreak
\subsection{The limits of ground based searches and the search for planets}

The range of lens masses which can give rise to observable lensing
events is very broad.  At the low end the practical limit is imposed 
by the finite size of the sources.  The amplification of a point
source which is perfectly aligned with a lens is infinite
(cf. eq. 11 with $ u = 0 $).  However, when a source with a finite
angular radius $ r_s $ is perfectly aligned, then its circular disk 
forms a ring--like image, which has its inner and outer radius, $ r_{in} $
and $ r_{out} $, given with a slightly modified eq. (9):
$$
r_{in} = \left[ \left( r_{_E}^2 + 0.25 r_s^2 \right) ^{1/2} - 0.5 r_s \right] ,
\hskip 2.0cm
r_{out} = \left[ \left( r_{_E}^2 + 0.25 r_s^2 \right) ^{1/2} + 0.5 r_s \right] ,
\eqno(50)
$$
where all quantities are expressed as angles.  Assuming, for simplicity,
a uniform surface brightness of the source the maximum magnification 
can be calculated as the ratio of the two areas:
$$
A_{max} = { \pi r_{out}^2 - \pi r_{in}^2 \over \pi r_s^2 } =
\left[ 4 \left( { r_{_E} \over r_s } \right) ^2 + 1 \right] ^{1/2}.  \eqno(51)
$$
For the event to be reasonably easy to detect we choose 
$ A_{max} \ge \sqrt{2} $,
which is equivalent to condition: $ r_s/r_{_E} \le 2 $.

An angular radius of a star is given as
$$
r_s = { R_s \over D_s } = 
2.3 \times 10^{-12} ~ rad 
~ \left( { R_s \over R_{\odot} } \right) ~ 
\left( { 10 ~ {\rm kpc} \over D_s } \right)
= 0.45 ~ \mu sec 
~ \left( { R_s \over R_{\odot} } \right) ~ 
\left( { 10 ~ {\rm kpc} \over D_s } \right) .
\eqno(52)
$$
This may be combined with the eq. (13) to obtain
$$
{ r_s \over r_{_E} } 
= 0.00050 ~ \left( { R_s \over R_{\odot} } \right) ~ 
\left( { M_{\odot} \over M } \right) ^{1/2}
\left( { 10 ~ {\rm kpc} \over D_s } \right) 
\left( { D_d \over 10 ~ {\rm kpc} } \right) ^{1/2}
\left( 1 - { D_d \over D_s } \right) ^{-1/2} .  \eqno(53)
$$
The smallest mass for which the Einstein ring radius is no less
than half the source radius is given as
$$
M \ge M_{min} \approx 6 \times 10^{-8} ~ M_{\odot} ~ 
\left( { R_s \over R_{\odot} } \right) ^2 ~
\left( { 10 ~ {\rm kpc} \over D_s } \right) 
\left( { D_d \over D_s - D_d } \right) .  \eqno(54)
$$

If the source is in the galactic bulge, at $ D_s \approx 8 $ kpc, and
the lens at $ D_d \approx 6 $ kpc (Zhao et al. 1995, 1996), then condition
(54) becomes:
$$
M_{min} \approx 2 \times 10^{-7} ~ M_{\odot} ~ 
\left( { R_s \over R_{\odot} } \right) ^2 ,
\hskip 2.0cm {\rm (galactic ~ bulge)} .  \eqno(55)
$$
If the source is in the LMC at $ D_s \sim 55 $ kpc, and the lens
is in the galactic halo at $ D_d \sim 10 $ kpc then the condition (54)
becomes:
$$
M_{min} \approx 1.0 \times 10^{-7} ~ M_{\odot} ~ 
\left( { R_s \over 7 ~ R_{\odot} } \right) ^2 ,
\hskip 2.0cm {\rm (LMC) } .  \eqno(56)
$$
Note that the scaling factor for the source radius is $ 7 ~ R_{\odot} $
in the eq. (56), while it is $ R_{\odot} $ in the eq. (55) to allow for the 
fact that the distance to the LMC is about 7 times larger than to the 
galactic center, and the source stars near the detection limit are 
correspondingly larger.  The time scale of a microlensing event 
caused by lenses at the lower mass
limit can be calculated combining the eqs. (15), (53) and (54) to obtain:
$$
t_0 \ge 30 ~ min ~ \left( { R_s \over R_{\odot} } \right) ~
\left( { D_d \over D_s } \right) ~
\left( { 200 ~ km ~ s^{-1} \over V } \right) .  \eqno(57)
$$

The estimate of $ M_{min} $ implies that it should be possible to extend
the searches down to masses as small as that of our moon.
These objects may be planets lensing together with their parent stars,
as described in section 4, or these may be planetary
mass objects roaming the interstellar space,
rather than orbiting any star.  Clearly, the task of finding them will
be very difficult unless they are very numerous.  Let us make a modest
assumption that every star has just one earth mass planet, and a
typical mass of the lenses actually responsible for the events detected
towards the bulge and the LMC is $ \sim 0.3 ~ M_{\odot} $.  This implies
that the mass ratio is $ \sim 10^{-5} $, and hence the optical depth
to microlensing by those planets is 
$ \sim 3 \times 10^{-6} \times 10^{-5} = 3 \times 10^{-11} $
towards the galactic bulge and 
$ \sim 10^{-7} \times 10^{-5} = 10^{-12}$ towards the LMC.
These numbers may be somewhat larger as the cross section for planetary
microlensing is enhanced by the presence of a nearby star (cf. Figure 9).
It is clear that the searches aiming at finding such planets must be 
vastly more thorough than anything operating now.

Let us make a crude estimate of the most sensitive ground based search
for microlensing that could be done by simply making multiple copies
of the existing systems.  Let all stars that may be microlensed, perhaps
$ 2 \times 10^8 $ of them, be measured every few minutes to detect all
events with a time scale $ t_0 \ge 30 $ minutes.  Let there be 
$ \sim 3000 $ hours of clear observing per year combining all ground
based sites.  This translates into six events per year with 
$ t_0 \sim 1 $ hour if the optical depth is $ \tau_{\rm 1 ~ hour} \approx
10^{-11} $.  The good news is that there are enough stars in the sky
to detect a moderately large number of planetary microlensing events 
with the 1--meter class ground based telescopes if a mega--project
is carried out for a few years.  The bad news is that the
required data rate would be about 1000 times larger than the current
MACHO rate.  The price tag would certainly be very impressive, but
probably not in excess of a single space mission.

This analysis demonstrates that with a rather immodest increase of
current microlensing searches the dark objects down to
asteroid mass range could be detected even if their fractional contribution
to the galactic mass was as small as $ \sim 10^{-5} $.  It is also possible,
at least in principle, to extend the search to masses as large as
$ 10^5 - 10^6 ~ M_{\odot} $.  The events due to lenses that massive have
a duration of $ \sim 100 $ years, yet they may be detected, at least
in principle, with two very different methods.  First, a parallax effect
due to Earth orbital motion will introduce $ \sim 1\% $ modulation in
the light curve, with a period of one year (Gould 1992).  Second, 
very numerous stars which are normally below the detection threshold
will be occasionally microlensed by a very large factor.  According
to the eq. (11) the highly magnified source is within factor of 2
of its peak brightness for a time interval 
$ t_{1/2} = t_0 \sqrt{3} / A_{max} $, which may be reasonably short for 
$ A_{max} \ge 100 $ (Paczy\'nski 1995b).  So, either of the two
methods may allow the extension of the microlensing searches to
the range of supermassive MACHOs.

\section{Acknowledgments}

It is a great pleasure to acknowledge the discussions with, and comments
by Drs. C. Alcock, D. Bennett, J. Ka\l u\.zny, A. Kruszewski, M. Kubiak, 
S. Mao, J. Miralda-Escud{\`e}, M. Pratt, K. Sahu, K. Z. Stanek, C. Stubbs, 
M. Szyma\'nski, A. Udalski, and J. Wambsganss.
This work was supported by the NSF grants AST-9216494 and AST-9313620.

{}

\vfill
\end{document}